\documentclass[aps,pra,twocolumn,superscriptaddress]{revtex4}
\usepackage[usenames,dvipsnames]{color}

\usepackage[latin1]{inputenc}
\usepackage{cancel}
\usepackage{graphicx}
\usepackage{amssymb}
\usepackage{braket,mleftright}
\usepackage{empheq}
\usepackage{subfigure}
\usepackage{stackrel}
\usepackage{soul}
\usepackage{blkarray}
\usepackage{multirow}
\usepackage{amsmath}
\usepackage{physics}
\usepackage{amsfonts}
\usepackage{bm}
\usepackage{bbold} 
\usepackage{color}
\usepackage[pdftex,colorlinks=true,urlcolor= blue,linkcolor=Blue,citecolor=RedViolet]{hyperref}
\newcommand{\beq}{\begin{equation}}
\newcommand{\eeq}{\end{equation}}
\newcommand{\beqq}{\begin{equation*}}
\newcommand{\eeqq}{\end{equation*}}
\newcommand{\bse}{\begin{subequations}}
\newcommand{\ese}{\end{subequations}}
\newcommand{\bea}{\begin{eqnarray}}
\newcommand{\eea}{\end{eqnarray}}
\newcommand{\beaa}{\begin{eqnarray*}}
\newcommand{\eeaa}{\end{eqnarray*}}

\begin{document}
\title{Unified framework to determine Gaussian states in continuous variable systems.}
\author{Fernando Nicacio}
\email{nicacio@if.ufrj.br} 
\affiliation{Instituto de F\'{\i}sica, Universidade Federal do Rio de
Janeiro, Caixa Postal 68528, Rio de Janeiro, RJ 21941-972, Brazil}
\author{Andrea Vald\'{e}s-Hern\'{a}ndez}
\affiliation{Instituto de F\'{\i}sica, Universidad Nacional Aut\'{o}noma de M\'{e}xico,
Apartado Postal 20-364, M\'{e}xico, Distrito Federal, México}
\author{Ana P. Majtey}
\affiliation{Facultad de Matem\'{a}tica, Astronom\'{\i}a y F\'{\i}sica, 
Universidad Nacional de C\'{o}rdoba,
Av. Medina Allende s/n, Ciudad Universitaria, X5000HUA C\'{o}rdoba, Argentina}
\affiliation{Consejo de Investigaciones Cient\'{\i}ficas y T\'{e}cnicas de la Rep\'{u}blica Argentina,
Av. Rivadavia 1917, C1033AAJ, Ciudad Aut\'{o}noma de Buenos Aires, Argentina}
\author{Fabricio Toscano}
\affiliation{Instituto de F\'{\i}sica, Universidade Federal do Rio de
Janeiro, Caixa Postal 68528, Rio de Janeiro, RJ 21941-972, Brazil}

\date{\today}
\begin{abstract}
Gaussian states are the backbone of quantum information protocols with continuous variable systems, 
whose power relies fundamentally on the entanglement between the different modes.  
In the case of global pure states, 
knowledge of the reduced states in a given bipartition of a multipartite quantum 
system bears information on the entanglement in such bipartition. 
For Gaussian states, the reduced states are also Gaussian, 
so there determination requires essentially the experimental determination of their covariance matrix. 
Here, we develop strategies to determine the covariance matrix of an arbitrary 
$n-$mode bosonic Gaussian state through measurement of the total phase acquired 
when appropriate metaplectic evolutions, associated with quadratic Hamiltonians, are applied. 
Simply one-mode metaplectic evolutions, such rotations, squeezing and shear transformations, 
in addition to a single two-mode rotation, allows to determine
all the covariance matrix elements of a $n-$mode bosonic system. 
All the single-mode metaplectic evolutions are applied conditionally to a state in which an ancilla qubit is entangled with the $n$-mode system. 
The ancillary system provides, after measurement, the value of the total phase of each evolution. The proposed method is experimentally 
friendly to be implemented in the most currently used continuous variable systems.

\end{abstract}
\maketitle

\section{Introduction}  
Quantum information processing is a research area devoted to study the information processing with quantum states. Its importance relies on the great advantages
its protocols have in comparison with the currently known protocols of classical information processing. The theoretical realm of quantum information processing comprises 
quantum computation or simulation, and quantum communication protocols, with emphasis on quantum teleportation and quantum cryptography. 
\par
Whereas the first advances regarding the theoretical development and the experimental implementation 
of quantum information processing arose in systems with finite Hilbert spaces,
more recently almost all the quantum information protocols have been extended to systems with infinite-dimensional Hilbert spaces, called continuous variable (CV) systems \cite{Braustein-book2003,braunstein05,Cerf-book2007,Andersen2010}.  
For example, the two schemes of quantum computation in discrete variables, quantum computation based on sequential applications of quantum gates, and the 
``one way'', measurement-based, quantum computation based on cluster states, were recently 
generalized to continuous-variable systems \cite{Lloyd1999,Menicucci2008,Menicucci2006}.
The same goes for the protocols of quantum teleportation, quantum cloning, quantum dense coding and quantum cryptography
\cite{Braunstein1998,Pirandola2006,Weedbrook2012,reid00,braunstein05}, some of which have already been experimentally implemented 
\cite{Furusawa1998,Yukawa2008-2,Miwa2009,Jouguet2013}.
\par 
It is worth noting that in the transition from discrete to continuous variable systems some advantages are gained, since several quantum information protocols are optimized using infinite-dimensional Hilbert spaces \cite{braunstein05}.
Moreover, the entanglement --the main resource in quantum processing protocols--, can be efficiently produced using squeezed light and linear optics \cite{WM94}. 
Besides, entanglement can also be detected more efficiently because the detectors for CVs in the optical domain are traditionally more efficient.
Indeed, the generation and manipulation of highly entangled states is achievable in CV systems \cite{chen14,yokoyama13}, and very often continuous-variable entanglement surpasses its discrete counterpart.  
\par 
When dealing with entangled CV systems, Gaussian states (GS) stand out as the paradigmatic ones  \cite{littlejohn1986,Schumake1986,Simon1987,Dutta1995,Eisert2003,Adesso2014}. These states 
constitute a powerful setting for 
quantum communication and quantum information protocols 
\cite{ferraro2005,Wang2007,Weedbrook2012}, and lie at the heart of 
CV optical and atomic technologies \cite{haroche-book,Leibfried2003,
Walborn-report2010}. 
Considerable effort has been devoted to characterizing the informational properties 
and the entanglement structure of GS \cite{simon00,Giedke2001,Werner2001,Eisert2003,Botero2003,Serafini2004,Adesso2004,Adesso2004-2,Adesso2005,Serafini2005,Adesso2007,Serafini2007,Marian2008,Adesso2007-2,Adesso2014,Cerf-book2007}. 
Particularly noteworthy is the exceptional role of GSs in
CV systems, since they are extremal with respect to  
various applications \cite{Wolf2006}.
\par
Highly multipartite entangled GSs (cluster states) can be produced, for example, 
by multimode frequency combs generated by 
a synchronously pumped optical parametric oscillator (SPOPO) \cite{Valcarcel2006,Menicucci2008,chen14,MedeirosdeAraujo2014,Gerke2015}.
Within this setup, a frequency comb with 60 entangled modes of the electromagnetic field was reported in \cite{chen14}. 
Also, Gaussian states are easy to prepare and control in trapped ions, atomic ensembles,
and opto-mechanical systems \cite{Cerf-book2007}. 
In particular, trapped ionic systems manipulated by laser light 
is now one of the most developed settings for the experimental investigation of quantum effects and 
processing of quantum information \cite{Cirac1995,Leibfried2003,Blatt2008}.  
The trapping potential confines the system to a harmonic motion in the 
vibrational modes, whose ground state is a GS \cite{Leibfried2003,Schuch2006}. 
A scheme of quantum computation over the vibrational modes of a single trapped
ion was recently suggested \cite{OrtizGutirrez2017}.
Highly entangled Gaussian
states can also be generated, to a good approximation, with twin photons generated in the spontaneous parametric down conversion (SPDC), since 
they can be performed as generalizations of two-mode squeezed states \cite{Walborn-report2010,tasca09a,tasca11,abouraddy07}. 
\par
The complete determination of a generic (non-Gaussian) state relies on a fully tomographic process. However,  GS
characterization is achieved by specifying only its first and second canonical moments.
The first moments can be freely adjusted by local phase-space displacements, 
and play no role in determining entanglement properties of the state. 
Instead, the second moments determine the so-called covariance matrix (CM), 
and fully characterize the relevant informational properties of the GS, 
particularly its entanglement structure.
\par
Each physical type of continuous variable systems has its particular method for the determination of the covariance matrix of the system state. 
In the context of quadrature modes of the quantized electromagnetic field the traditional method is homodyne detection, which
involves the interference of the input field to be 
probed with a local oscillator in a beam splitter. In this case, the value of the chosen measured quadrature is directly obtained from the difference of the photo-current at the output ports of the beam splitter.
The fluctuations around the mean values give the variances
needed to infer the matrix elements of the CM of the input field quantum state.
In CV systems where the quadrature measurement is not directly accessible,
there exist two different strategies to determine the covariance matrix. These CV systems
are generically massive oscillators and the first strategy
involves the measurement of a qubit ancilla properly coupled to the oscillators,
which directly gives phase-space values of the Weyl characteristic function of the quantum state.
For one-vibrational modes of a trapped ion this strategy was outlined in
\cite{Wallentowitz1995}, and later generalized for a network of oscillators in \cite{Tufarelli2012}. The second strategy is more suitable 
in the context of optomechanical systems. It consists in using a CV probe entangled with the oscillators
\cite{Vanner2014,Moore2016}. In this case an intracavity electromagnetic mode is coupled via radiation pressure 
to a mechanical mode through one mobile cavity mirror. 
The covariance matrix of the mechanical mode is inferred 
through homodyne detection of the leaking field of the cavity, which contains information about the intracavity mode and hence about the mechanical mode.
In the context of the CV system corresponding to the spatial transverse modes of single
photons, the best method available to determine the CM of a quantum state (Gaussian or not) was reported in \cite{hormeyll14}.

Here we present a unified method to determine the covariance matrix of Gaussian states that can be implemented in any CV system.
The tools involved in our method are unitary evolutions that preserve the Gaussianity 
of the evolved state, and the total phase acquired by the state under such evolutions. 
The former corresponds to the metaplectic group of unitary operations, $\hat M_{\sf S}$,
generated by quadratic Hamiltonians in the position and momentum canonical conjugate operators \cite{Dutta1995,gossonbook}, which are characterized by a symplectic matrix ${\sf S}$ 
\cite{gossonbook}.   
The second tool is 
the total phase acquired by the Gaussian state $\hat \rho_G$ through the evolution, given by \cite{Khoury2014}:   
\begin{eqnarray}
\phi=\arg\left[{\rm Tr} ( \hat M_{\sf S} \hat \rho_G)\right].
\end{eqnarray}
This is a particular case of a general extension of the total phase $\phi=\arg\left[\expval{\hat U}{\psi}\right]$, originally defined for pure states 
$\hat \rho=\dyad{\psi}{\psi}$, where it was defined as the sum 
 of the geometric and dynamical phases of the evolution \cite{mukunda1993}.
 \par
The feasibility of the method developed here relies basically on two main features. The first one is that the required unitary evolutions are one-mode metaplectic operators 
(such as rotations, shearings and squeezings 
\cite{littlejohn1986,ozorio1998,gossonbook}), and a single two-mode rotation
(i.e., a beam-splitter like rotation \cite{Dutta1995}). 
The second one is that these evolutions imprint the information of the 
covariance matrix elements in the corresponding total phases $\phi$. Hence, by means of an experimentally friendly protocol for measuring these total phases, the information of the full covariance matrix can be recovered, irrespective of the CV system involved. In particular, here we propose such experimental protocols in three paradigmatic CV systems: 
the quantized electromagnetic field, the vibrational modes in trapped ions, 
and the transverse spatial degrees of freedom of entangled single photons.
\par
The work is structured as follows. 
In Section \ref{II} we review the Weyl-Wigner formalism 
that will allow us to calculate the total phase acquired by 
Gaussian states under arbitrary metaplectic evolutions. 
The metaplectic group, with special attention paid to the Weyl and Wigner symbols of the metaplectic operators, 
is introduced in Section \ref{III}. 
The total phase acquired by an $n-$mode arbitrary GS under metaplectic evolutions 
is calculated in Section \ref{IV}. 
In Section \ref{V} we present the strategies that allow for full determination of the covariance matrix (and hence the GS) 
through the implementation of appropriate metaplectic evolutions in different copies of the GS, plus further measurement of the corresponding acquired phases. Section \ref{VI} is devoted to a brief outline of the main features of gaussian entanglement, and the applicability of our method to determine entanglement in bipartitions having $1\times(n-1)$ modes in GS. 
The general experimental protocol aimed at measuring the total phase acquired by a general state under an arbitrary unitary evolution is described in Section \ref{VII}.
We also discuss specifically the case of metaplectic operations over GS.
In Section
\ref{VIII} we describe the implementation of the protocol in the context of the
spatial degrees of freedom of twin photons 
generated in the SPDC, trapped ions, and
quantized modes of the electromagnetic field. 
Finally, some conclusions and final remarks are provided in Section \ref{Conclusions}.

\section{Weyl-Wigner Formalism}
\label{II}
 \par
We consider a multipartite system composed of $n$ bosonic modes, 
described through the column vector of operators denoted by 
$\hat x := (\hat q_1,\hat p_1,...,\hat q_n,\hat p_n){\!^\top}$, 
where $\hat q_j$ and $\hat p_j$ stand for the position and momentum operators, 
respectively, of the $j^{\underline{\text{th}}}$ mode. 
The usual commutation-relation between these operators can be succinctly written as 
$[\hat x_j,\hat x_k] = i \hbar \mathsf J_{jk}$, where $\hat{x}_{j(k)} $ is the $j^{\underline{\text{th}}} $ $(k^{\underline{\text{th}}})$ component of $\hat x$, and
$\mathsf J_{jk}$ being the elements of the $2n \times 2n$ symplectic matrix 
\begin{equation}                                                                         \label{SimpJ}
\mathsf J = \bigoplus_{j=1}^n\;{\sf J}_2 \;\;,\;\;
{\sf J}_2 \equiv  \begin{pmatrix} 
      0 & 1 \\
     -1 & 0 \\
   \end{pmatrix},
\end{equation}
such that ${\sf J}^\top=-{\sf J}={\sf J}^{-1}$.
When dealing with only one mode, it is useful to define the two component co\-lumn 
vector of canonically conjugate operators 
$\hat x^{(j)} = (\hat q_j,\hat p_j){\!^\top}$ such that 
$[\hat x^{(j)}_k,\hat x^{(j)}_l] = i \hbar ({{\sf J}_2})_{kl}$.  
\par
An alternative description of $n-$bosonic modes, more often used in the context of second quantization, 
resorts to the annihilation and creation 
operators of each mode, 
$\hat a_{j}:=(1/\sqrt{2\hbar})(\hat q_j+i\hat p_j)$, and 
$\hat a_{j}^\dagger:=(1/\sqrt{2\hbar})(\hat q_j-i\hat p_j)$, respectively. 
These operators satisfy the bosonic commutation relations $[\hat a_j,\hat a_k^\dagger]=\delta_{jk}$,
and $[\hat a_j,\hat a_k]=[\hat a^\dagger_j,\hat a^\dagger_k]=0$.
We will resort to this description only to specify some quadratic Hamiltonians 
that will appear in the next sections.
\par
The Weyl translation operator is defined as \cite{littlejohn1986,ozorio1998,gossonbook}
\begin{equation}                                                                         \label{HeisOp}
\hat T_\xi:= \hat T_{\xi^{(1)}}\otimes\ldots\otimes\hat T_{\xi^{(n)}} = 
             e^{\frac{i}{\hbar} \hat x^{\!\top} {\sf J}\xi },  
\end{equation}
where we define the \textit{chord} 
$\xi := (\xi_{q_1},\xi_{p_1},...,\xi_{q_n},\xi_{p_n}){\!^\top}$ 
and $\xi^{(j)} := (\xi_{q_j},\xi_{p_j})$. 
Note that the chord 
is a column vector
indicating the direction of the translation of the canonically conjugate operators, i.e.,  
$\hat T^\dag_{\xi}\hat x \hat T_{\xi}=\hat x+\xi\hat 1$.
Notice that the translation operator is unitary, so $\hat T_\xi^{-1}=\hat T_\xi^\dagger=\hat T_{-\xi}$.
\par
The symplectic Fourier transform of $\hat T_\xi$ is known as the 
reflection operator \cite{Royer1977,ozorio1998}, namely
\bea                                                                       \label{ReflOp}
\hat R_x &:=&\frac{1}{2^n}\mathbb{\hat R}_x = 
\frac{1}{2^n}(\mathbb{\hat R}_{x^{(1)}}\otimes\ldots\otimes \mathbb{\hat R}_{x^{(n)}})=\nonumber\\
         &=&  \int \frac{d\xi}{(4\pi\hbar)^n} \,  
                          e^{\frac{i}{\hbar}  \xi^{\!\top} {\sf J}x}\, \hat T_\xi,  
\eea
which is an Hermitian and unitary (hence involutory) operator, that is, $\hat R_x^2=\hat 1$.
Here the \textit{center} $x := (q_1,p_1,...,q_n,p_n)^\top$ is a column vector in phase space indicating 
the reflection point, i.e., $\hat R_x \hat x \hat R_x=-\hat x+2x\hat 1$. We also define 
$x^{(j)} := (q_j,p_j)$.
\par
An arbitrary operator $\hat A$ acting on the Hilbert space of the continuous-variable ($n$-mode) system, can be uniquely expanded as a linear combination of either translation (\ref{HeisOp}) or reflection (\ref{ReflOp}) operators \cite{ozorio1998,gossonbook}. These expansions constitute, respectively, the Weyl and the Wigner representation 
of $\hat A$: 
\begin{subequations}                                                                     \label{HWrep}
\begin{eqnarray}                                                                         
\hat A &=&  \int \frac{d\xi }{(2\pi\hbar)^n}\!  \, \mathcal{A}(\xi) \,  \hat T_\xi,               \label{HWrep-a}\\ 
\hat A &=& \int \frac{d x}{(2\pi\hbar)^n}  \, A(x) \, \mathbb{\hat R}_x.                \label{HWrep-b}
\end{eqnarray}
\end{subequations}
The coefficients $\mathcal{A}(\xi)$ and $A(x)$ are, respectively, 
the Weyl and the Wigner symbols of the operator $\hat A$, 
given by 
\begin{equation}                                                                         \label{HWsymb}
\mathcal{A}(\xi) = \textrm{Tr}\,(\hat A \hat T_{\xi}^\dag), \,\,\,\,     
A(x)   = \textrm{Tr}\,( \hat A \mathbb{\hat R}_x),
\end{equation}
by virtue of \cite{ozorio1998}
\bse  \label{trace}
\begin{eqnarray}                                                                         \label{trace-translation}
\textrm{Tr}\,(\hat T_\xi \hat T^ \dag_{\xi'}) &=& (2\pi\hbar)^n\delta(\xi'-\xi) ,                  \nonumber \\
\textrm{Tr}\,(\mathbb{\hat R}_{x}\mathbb{\hat R}_{x'}) &=& (2\pi\hbar)^n\delta(x'-x).
 \label{trace-reflection}
\end{eqnarray}
\ese
The Weyl and Wigner symbols are related to each other via a symplectic Fourier transform, 
{\it viz.}, 
\bse                                                                                     \label{Weyl-Wigner-expansions}
\begin{eqnarray}                                                                         
\mathcal{A}(\xi)&=&  \int \frac{dx }{(2\pi\hbar)^n}\, A(x)\;
                        e^{\frac{i}{\hbar}  \xi{\!^\top} {\sf J} x},    \label{Wigner-expansion}           \\                       
A(x)&=&  \int \!  \frac{d\xi}{(2\pi\hbar)^n} \, \mathcal{A}(\xi)\;
                        e^{\frac{i}{\hbar}x^{\!\top} {\sf J} \xi }.
                         \label{Weyl-expansion}  
\end{eqnarray}
\ese
In particular, the Wigner function $W(x)$ of a quantum state is (a normalized version of) 
the Wigner symbol associated with the corresponding density operator $\hat \rho$ \cite{Grossmann1976,Royer1977,ozorio1998}, that is, 
\begin{equation}                                                                         \label{Wigfunc}
W(x) := \Tr \left[ \frac{\hat \rho}{(2\pi\hbar)^n} \mathbb{\hat R}_x \right].  
\end{equation}
Its symplectic Fourier transform is the Weyl symbol (or characteristic function)
of $\hat \rho$ \cite{ozorio1998}:
\begin{equation}                                                                         \label{Charfunc}
\chi(\xi) \! = \!\! \int \!\! \frac{d x}{(2\pi\hbar)^n} \, W(x) \,
                              e^{ \frac{i}{\hbar} \xi^{\!\top}{\sf J} x }
           =   \Tr \left[ \frac{\hat \rho}{(2\pi\hbar)^n} \hat T^{\dagger}_{\xi} \right].  
\end{equation}
Thus, for example, for a Gaussian state with null mean values, Eqs.~(\ref{Wigfunc}) and (\ref{Charfunc}) lead to 
\begin{equation}                                                                         \label{WigGauss}
W_{\textit{G}}(x) = \frac{1}{(2\pi\hbar)^n} \frac{ \exp \! \left[ -\frac{1}{2\hbar} x^{\!\top} {\bf V}^{-1} x  \right]}
             { \sqrt{\det \bf V} },  
\end{equation}
and
\begin{equation}                                                                         \label{CharGauss}
\chi_{\textit{G}}(\xi) =  \frac{1}{(2\pi\hbar)^n}  
\exp \left[ -\frac{1}{2\hbar}  \xi^{\!\top} \mathsf J^\top {\bf V}\mathsf J \xi \right],  
\end{equation}
where $\bf V$ is the $2n \times 2n$ covariance matrix 
with elements  
\begin{equation}                                                                         \label{CMdef}
{\bf V}_{ij} = \frac{1}{2\hbar} 
{\rm Tr} \left[ \hat \rho \left(\hat x_i \hat x_j + \hat x_j \hat x_i\right) \right].  
\end{equation}
Notice that since the mean values of a general state $\hat \rho$ 
can be made equal to zero by simply performing a translation 
[that is, a local operation in each mode, see Eq.(\ref{HeisOp})]
according to $\hat T^{\dagger}_{\xi}\hat \rho \hat T_{\xi}$, 
with $\xi=-\Tr(\hat \rho \hat x)=-(\Tr(\hat \rho \hat x_1),\ldots,\Tr(\hat \rho \hat x_{2n}))^\top$,
we can restrict our attention to Gaussian states with null mean values, without loss of generality.

\section{Metaplectic Group and its Weyl-Wigner representations}     
\label{III}
In this Section we introduce the metaplectic operators associated with unitary evolutions 
under quadratic Hamiltonians, and focus on their corresponding 
Weyl and Wigner symbols. These symbols will allow us to calculate the 
total phase corresponding to metaplectic evolutions over Gaussian states
in the further coming.
\par
Quadratic Hamiltonians are defined as those of the form 
\begin{equation}                                                                         \label{Quadham}
\hat H = \frac{\omega}{2}\hat x^{\!\top} {\bf H} \hat x, 
\end{equation}
where ${\bf H}$ is a $2n \times 2n$ symmetric real matrix 
known as the Hessian of $\hat H$, and $\omega$ is a real parameter.
These Hamiltonians constitute the algebra $\mathfrak{mp}(2n,\mathbb R)$
of the
Metaplectic group \cite{gossonbook,Dutta1995,littlejohn1986,ozorio1998}. 
As is usual for Lie groups, when exponentiating elements in the algebra
we obtain elements of the group,
\begin{equation}                                                                         \label{Metap}
\hat M_{\mathsf S} := e^{-\frac{i\omega t}{2\hbar} \hat x^{\!\top} {\bf H} \hat x}.   
\end{equation}
Here the subindex $\mathsf S$ highlights the relation between the metaplectic operator $\hat M_{\mathsf S}$ in Eq.~(\ref{Metap})
and the matrix 
\begin{equation}                                                                         \label{Symp}
\mathsf S := e^{\mathsf J {\bf H} \omega t},   
\end{equation}
which is an element of the real symplectic group ${\rm Sp}(2n,\mathbb R)$,
defined as the set of matrices such that 
$\mathsf S^{\top} \mathsf J \mathsf S = \mathsf J$. 

Note that $\mathsf J {\bf H} $  is an element of the symplectic algebra
$\mathfrak{sp}(2n,\mathbb R)$ that is in one-to-one correspondence with the 
element in Eq.~(\ref{Quadham}), that belongs to the algebra $\mathfrak{mp}(2n,\mathbb R)$. However, it may be that for some matrices ${\sf J} {\bf H}$ 
there are two values of  $\omega t$ that give the same symplectic matrix 
${\sf S}$ in Eq.~(\ref{Symp}).
This is a manifestation of the fact that the metaplectic group is a double covering group of 
the symplectic one, i.e., there are two metaplectic operators, namely 
$\pm \hat M_{\mathsf S}$, 
associated with each symplectic matrix $\mathsf S$ \cite{Dutta1995,littlejohn1986}. 
Another peculiar characteristic of the metaplectic group is that, as occurs in the 
symplectic group, it is not an exponential group
\cite{Dutta1995}. Thus, there are elements in 
${\rm Mp}(2n,\mathbb R)$, as in ${\rm Sp}(2n,\mathbb R)$, that cannot be written as an exponentiation of some element
in $\mathfrak{mp}(2n,\mathbb R)$, and $\mathfrak{sp}(2n,\mathbb R)$ respectively, but rather 
decompose into products of operators like that in 
Eq.~(\ref{Metap}). In this case the associated symplectic matrix is the product of symplectic matrices like 
those in Eq.~(\ref{Symp}), corresponding to each factor of the metaplectic decomposition.
In fact, any symplectic matrix can be written as a product of another symplectic matrices 
in a non unique way. This leads to different decompositions for the associated 
metaplectic operator.
In particular, 
it will be useful for latter purposes to resort to the factorization proved in \cite{gossonbook} that establishes 
that every ${\sf S}\in {\rm Sp}(2n,\mathbb R)$ can be written as
\beq
\label{decom-S}
{\sf S}={\sf S}^{\prime}{\sf S}^{\prime\prime},
\eeq
where ${\sf S}^\prime$ and ${\sf S}^{\prime\prime}$ are symplectic matrices that are products 
of matrices of the form (\ref{Symp}), and such that $\det({\sf S}^{\prime}+{\sf I}_{2n})\neq 0$ and 
$\det({\sf S}^{\prime\prime}+{\sf I}_{2n})\neq 0$ (that is, neither ${\sf S}^\prime$ nor ${\sf S}^{\prime\prime}$ has an eigenvalue equal to $-1$). The metaplectic operator corresponding to ${\sf S}$ as given by Eq.~(\ref{decom-S}) 
can be chosen as

\beq
\label{decom-M}
\hat M_{\mathsf S} = \pm \hat M_{{\sf S}^{\prime}} \hat M_{{\sf S}^{\prime\prime}}.
\eeq
The indeterminacy of the signal $\pm$ is removed once we specify 
the time dependence of the symplectic matrix ${\sf S}={\sf S}(t)$ as we will see
in what follows.

The Weyl and Wigner symbols of the metaplectic operator (\ref{Metap})  
are given, respectively, by \cite{mehlig2001, gossonbook}
\begin{equation}                                                                         \label{Hsmet}
\mathcal{M}_\mathsf{S}(\xi) = 
\frac{i^{\nu^{-}_{\mathsf S}}
\exp\left[
-\frac{i}{4\hbar} \xi^{\!\top}
\mathsf{J}\mathbf{C}^{^{^{\!\!\!{-\!1}}}}\!\!\!_{\mathsf S}
\mathsf{J}\xi\right]}
{
\sqrt{\left|\det\left(\mathsf S - \mathsf I_{2n}\right)\right|}},                                                          
\end{equation}
and
\begin{equation}                                                                         \label{Wsmet}   
M_\mathsf{S}(x) =  
\frac{2^n i^{\nu^{+}_{\mathsf S}}
\exp\left[-\frac{i}{\hbar} x^{\!\top} \mathbf{C}_{\mathsf S}x\right]}
{\sqrt{\left|\det\left(\mathsf S + \mathsf I_{2n}\right)\right|}}.           
\end{equation}
Here the symmetric matrix

\begin{equation}                                                                         \label{Cayley1}
{\bf C}_\mathsf{S} =
-\mathsf{J} \frac{\left(\mathsf{S}- \mathsf{I}_{2n} \right)}
           {\left(\mathsf{S}+ \mathsf{I}_{2n}  \right)}
\end{equation}
stands for the {\it Cayley parametrization} of $\mathsf S$. 
Note that, depending on $\mathsf S$, 
the above symbols may not be defined, 
since ${\bf C}_\mathsf{S}$ or its inverse may not exist. When both symbols $\mathcal{M}_\mathsf{S}(\xi)$ and $M_\mathsf{S}(x)$ in Eqs.~(\ref{Hsmet}) and (\ref{Wsmet}) have no divergencies, they are related by the symplectic Fourier
transform, and the index $\nu^{+}_{\mathsf S}$ is given by
\begin{equation}                                                                         \label{CZdef}
\nu^{+}_{\mathsf S} = \nu^{-}_{\mathsf S} + 
                      \tfrac{1}{2} {\rm Sng}\, {\mathbf C}_{\mathsf S} 
                      \,\, ({\rm mod} \, 4), 
\end{equation}
where ${\rm Sng} \, \bf X$ is the number of positive eigenvalues minus 
the number of negative eigenvalues of the matrix $\bf X$, and $\nu^{-}_{\mathsf S}$ is the {\it Conley-Zehnder} (CZ) index \cite{Conley1984,ozorio1998,mehlig2001,gossonbook}. 
This index determines the sign of the metaplectic operator associated with the single matrix $\mathsf S$. This can be summarized in the definition: 
\begin{equation}
\sqrt{\det\left(\mathsf S - \mathsf I_{2n}\right)} := 
 i^{ - \nu^{-}_{\mathsf S}} 
\sqrt{\left|\det\left(\mathsf S - \mathsf I_{2n}\right)\right|},
\end{equation}
where $\nu^{-}_{\mathsf S}$ 
acquire the values $\{0,2\}$ if $\det\left(\mathsf S - \mathsf I_{2n}\right)>0$,
and $\{1,3\}$ if $\det\left(\mathsf S - \mathsf I_{2n}\right)<0$. 
For an invertible ${\bf C}_{\mathsf S}$, 
$\tfrac{1}{2}{\rm Sng} \, {\bf C}_{\mathsf S}$ is an integer, 
thus $\nu^{+}_{\mathsf S}$ is also an integer number in the set
$\{0,1,2,3\}$ \cite{gossonbook}, in accord with Eq.~(\ref{CZdef}).
Notice that the symbols in (\ref{Hsmet}) and in (\ref{Wsmet}) diverge, respectively, 
when an eigenvalue of $\mathsf S$ becomes $1$ and $-1$. In this case the symbols do exist, yet they are not calculated via Eqs.~(\ref{Hsmet}) and (\ref{Wsmet}), but instead using for example, Eq.~(\ref{compo-law}).
\par
Here we are interested in metaplectic operators associated with a temporal evolution, 
so let us assume that $\mathsf S$ depends continuously on a real parameter $t$. 
An example is given in (\ref{Symp}), where $\sf S$ belongs to a uniparametric subgroup
of ${\rm Sp}(2n,\mathbb R)$; however, in the general case ${\sf S}={\sf S}(t)$
does not necessarily belong to any uniparametric subgroup.
At each time $t$ the metaplectic operator 
associated with ${\sf S}$ has a definite sign that can be traced out 
by continuity of the operator with respect to $\mathsf S$, in accord with 
\begin{equation}                                                                         \label{CZlimit}
\lim_{t \to 0^+} \mathsf S = {\sf I}_{2n} \Longrightarrow
\lim_{t \to 0^+} \hat M_{\mathsf S}  =  + \hat{\sf 1}.
\end{equation}
This continuity property reflects in 
the behavior of the Weyl and Wigner symbols of $\hat M_{\mathsf S}$
through the indexes  $\nu^{\pm}_{\mathsf S}$, which
must change accordingly whenever there exists a discontinuity of the symbol $\mathcal{M}_{\mathsf S}(\xi)$ or $M_{\mathsf S}(x)$ in Eqs.(\ref{Hsmet}) and in (\ref{Wsmet}), that is, whenever 
${\sf S}$ has an eigenvalue $1$ or $-1$, respectively. 
For example, when $t=0$ the Wigner symbol of the identity operator is 
${\sf 1}(x)=1$, so $\nu^+_{\mathsf S} = 0$ for $t=0$ and all $t>0$ until 
an eigenvalue of ${\sf S}$ becomes $-1$, which occurs, say, at $t=t^*$. Then, as long as $\mathsf S(t^*)$ does not have an eigenvalue equal to $1$, we can switch the representation and calculate the 
Weyl symbol of the metaplectic operator. 
The continuity 
of the symbols in the vicinity of $t=t^*$ is guaranteed by the relation in Eq.~(\ref{CZdef}). 
If at some time $t$, ${\mathsf S(t)}$ has simultaneous eigenvalues, $1$ and $-1$, we rely on the decomposition in Eq.~(\ref{decom-M}) and calculate 
the Wigner symbol of the composition such that $ \hat M_{\mathsf S} =\hat M_{{\sf S}^\prime} \hat M_{{\sf S}^{\prime\prime}} $ 
using the following expression \cite{ozorio1998}:
\bea
M_{\mathsf S}(x)&=&
\int \frac{dx'}{(\pi\hbar)^n}\int \frac{dx''}{(\pi\hbar)^n}
M_{{\sf S}^\prime}(x')M_{{\sf S}^{\prime\prime}}(x'')\times \nonumber\\
&&\times e^{\frac{2i}{\hbar}(x''-x)^\top {\sf J}(x'-x)},
\label{compo-law}
\eea
where $M_{{\sf S}^\prime}(x)$ and $M_{{\sf S}^{\prime\prime}}(x)$ are the 
Wigner symbols of $\hat M_{{\sf S}^\prime}$ and $\hat M_{{\sf S}^{\prime\prime}}$
respectively, whose structure is given in Eq.~(\ref{Wsmet}).
%

\section{Total Phase of Gaussian states under Metaplectic Evolutions}     
\label{IV}
\par
Consider the unitary evolution $\hat U = e^{-\frac{i}{\hbar} \hat H t}$ generated by the Hamiltonian $\hat H$. As a quantum state $\hat \rho$ evolves accordingly, it acquires a total phase defined as \cite{mukunda1993} 
\begin{equation}                                                                         \label{Totpha}
\phi=\arg\left[{\rm Tr} ( \hat U \hat \rho )\right], 
\end{equation}
with the argument function defined as
\begin{equation}
\arg(x+iy) = \left\{ \begin{array}{ll}
                      \arctan(\tfrac{y}{x})       & \text{if} \, x > 0 ; \\
                      \arctan(\tfrac{y}{x}) + \pi & \text{if} \, x < 0 \, \& \, y \ge 0 ; \\
                      \arctan(\tfrac{y}{x}) - \pi & \text{if} \, x < 0 \, \& \, y < 0 ;   \\ 
                      + \frac{\pi}{2}             & \text{if} \, x = 0 \, \& \,  y > 0 ;  \\
                      - \frac{\pi}{2}             & \text{if} \, x = 0 \, \& \,  y < 0 ;  \\
                      \text{undefined}            & \text{if} \, x = y = 0.               
                      \end{array}
             \right. 
\end{equation}
%
This implies that $-\pi\le\phi\le \pi$. It is important to notice, for future analysis, that we always have $\left|\Tr\left(\hat U\hat \rho\right)\right|\leq 1$. 

In what follows we will calculate this phase for an initial $n-$mode arbitrary Gaussian state with null mean value, 
$\hat \rho=\hat \rho_G$, subject to a unitary 
evolution generated by a generic metaplectic operator $\hat U=\hat M_{\mathsf S}(t)$, that is, a generic composition of evolutions 
like those in Eq.~(\ref{Metap}).
As for the density operator, we can expand it in the Weyl representation, 
with the coefficients given by Eq.~(\ref{CharGauss}), or rather we can resort to its Wigner representation, and employ Eq.~(\ref{WigGauss}). Then, using Eqs.~(\ref{HWrep}) for $\hat A=\hat M_{\mathsf S}$, we can write ${\rm Tr}( \hat \rho_{G} \, \hat M_{\mathsf S})$ as
\bse
\label{Arg1aa}
\begin{eqnarray}                                                                          
{\rm Tr}( \hat \rho_G \, \hat M_{\mathsf S} ) 
&=&  \int\!\!d\xi \, \chi_{\it{G}}(\xi) \,  \mathcal{M}_{\mathsf S}(-\xi)     \\
&=& \int\!\!dx \, W_{\it{G}}(x) M_{\mathsf S}(x).                                                                                       
\end{eqnarray}
\ese 
We now resort to Eqs.~(\ref{WigGauss}), (\ref{CharGauss}), (\ref{Hsmet}) and (\ref{Wsmet}), and perform the Gaussian integrations to get 
\bse
\begin{eqnarray}                                                                          
\!\!\!\!\!\!\!\!\!\!\!\!{\rm Tr}( \hat \rho_G \hat M_{\mathsf S})\! 
&=& \!\!\frac{ i^{\nu^-_{\mathsf S}} }
          { \!\sqrt{\left|\det\left(\mathsf S - \mathsf I_{2n}\right)\right| 
 \det\! \left( {\bf V} - \frac{i}{2} {\bf C}^{-1}_\mathsf S \right)}}\label{Arg1} \\                                       
 &=& \!
\frac{ i^{\nu^{+}_{\mathsf S}}}
{\sqrt{ \! \left| \det\!\left( \mathsf S + \mathsf I_{2n} \right) \right|
        \det(\frac{1}{2}{\mathsf I}_{2n} \!+\! i {\bf V}{\bf C}_\mathsf S ) }},                         \label{Arg2}                                                      
\end{eqnarray} \\
\ese \\
where $\sqrt{z}$ is the square root with a positive real part of the complex number $z$.
Notice that Eq.~(\ref{Arg1}) holds whenever $\mathsf S$ does not have an eigenvalue 
equal to $1$, whereas Eq.~(\ref{Arg2}) is valid as long as none of the eigenvalues of $\mathsf S$ equals $-1$. 
If $\mathsf S$ possess eigenvalues $1$ \textit{and} $-1$, we can resort to the factorization 
in Eq.~(\ref{decom-S}), decompose 
$ \hat M_{\mathsf S}$ according to Eq.~(\ref{decom-M}), and write
\bea
{\rm Tr} ( \hat \rho_G \hat M_{\mathsf S} )& =& 
{\rm Tr} (\hat \rho_G \hat M_{{\sf S}^\prime} \hat M_{{\sf S}^{\prime\prime}})\nonumber\\
&=&
\int\!\!dx \, W_{\it{G}}(x) M_{\mathsf S}(x),\label{TrComp}
\eea 
where the symbol $M_{\mathsf S}(x)$ is given in Eq.~(\ref{compo-law}). After performing the integration in (\ref{TrComp}), we obtain
\begin{widetext}
\begin{equation}
\label{Arg3}
{\rm Tr} (\hat \rho_G \hat M_{\mathsf S} ) = 
\frac{ i^{\nu^+_{\mathsf S^{\prime} } + \nu^+_{\mathsf S^{\prime\prime}} } }
     { \sqrt{  \left| \det\!\left( \mathsf S^{\prime} + \mathsf I_{2n} \right) 
                      \det\!\left( \mathsf S^{\prime\prime} + \mathsf I_{2n} \right) \right|}}
\frac{ \sqrt{\det({\bf V} + \tfrac{i}{2}\mathsf J)^{-1}} }
     {\sqrt{ \det \left[ ({\bf V} - \tfrac{i}{2}\mathsf J) -   
                                    ({\bf V} + \tfrac{i}{2}{\bf C}_{\mathsf S^{\prime\top}}) 
                                    ({\bf V} + \tfrac{i}{2}\mathsf J)^{-1}
                                    ({\bf V} + \tfrac{i}{2}{\bf C}_{\mathsf S^{\prime\prime\top}}) \right] }}.
\end{equation}
\end{widetext}
Finally, Eqs.~(\ref{Arg1}), (\ref{Arg2}), and (\ref{Arg3}), together with Eq.~(\ref{Totpha}), allow us to write the total phase $\phi_{\sf S}[\hat \rho_G]$ acquired by an
arbitrary (null mean value) $n-$mode Gaussian state evolving under the metaplectic evolution $\hat M_{\sf S}$ as 
\begin{widetext}
\bse
\label{total-phase-total-system}
\bea
\phi_{\sf S}[\hat \rho_G] &=& \frac{\pi}{2} \nu^-_{\mathsf S} - 
                  \frac{1}{2}{\rm arg}
                                     \left[
                                           \det \left( {\bf V} - 
                                           \tfrac{i}{2} {\bf C}^{-1}_\mathsf S \right)
                                    \right]                                            \label{TotalPha2} \\
              &=& \frac{\pi}{2} \nu^+_{\mathsf S} - 
                  \frac{1}{2}{\rm arg}\left[
                                           \det(\tfrac{1}{2}{\mathsf I}_{2n} + 
                                           i {\bf V}{\bf C}_\mathsf S ) 
                                     \right]                                           \label{TotalPha3}  \\
                                   &=& \frac{\pi}{2}({\nu^+_{\mathsf S^\prime} + 
                                                       \nu^+_{{\sf S}^{\prime\prime}}}) 
             - \frac{1}{2}{\rm arg}  
                  \left[\det \left[ ({\bf V} - \tfrac{i}{2}\mathsf J) -   
                                    ({\bf V} + \tfrac{i}{2}{\bf C}_{\mathsf S^{\prime\top}}) 
                                    ({\bf V} + \tfrac{i}{2}\mathsf J)^{-1}
                                    ({\bf V} + \tfrac{i}{2}{\bf C}_{\mathsf S^{\prime\prime\top}}) 
                  \right] \right].                                                      \label{TotalPha4}
\eea 
\ese
\end{widetext}
Here Eq.~(\ref{TotalPha2}) holds whenever $\mathsf S$ does not have an eigenvalue 
equal to $1$, Eq.~(\ref{TotalPha3}) whenever $\mathsf S$ does not have an eigenvalue 
equal to $-1$, and Eq.~(\ref{TotalPha4}) for any symplectic matrix $\sf S$ once the factorization in Eq.~(\ref{decom-S}) is found. 
For a matrix $\mathsf S$ that does not have eigenvalues $\pm 1$, the three equations above coincide.
\par
\section{Determination of the Gaussian state through the total phase }
\label{V}
In what follows we exploit Eq.~(\ref{TotalPha3}) to design strategies that allow for the complete determination of the covariance matrix ${\bf V}$ ---hence to completely specify 
an arbitrary Gaussian state $\hat \rho_G$--- by the mere implementation of appropriate metaplectic 
evolutions over one and two modes, once an adequate measure of the total phase acquired in each evolution is performed. 
This Section deals specifically with the development of the strategies (assuming that the total phases are known), whereas a particular experimental protocol 
for measuring such phases in the context of any CV system is left for Section \ref{VI}. 

Since the state $\hat \rho_G$ to be determined will evolve under suitable metaplectic evolutions, we will refer to it as \textit{initial state}. 
In addition, $\hat \rho_G$ is a quantum state in the interaction picture representation, in relation to a free evolution of the CV system represented by 
a metaplectic evolution (generically, though not necessarily, an harmonic one). Therefore, the total phase $\phi$ is the phase of the evolution in the 
interaction picture, associated exactly with the Hamiltonian in Eq.~(\ref{Quadham}), without any constant factors added.

We start by writing the $n-$mode covariance matrix of the initial state $\hat \rho_G$ in the block form 
\begin{widetext}
\begin{equation}                                                                         \label{V-block}
{\bf V} = \begin{pmatrix}
{\bf V}^{(1)}&{\bf E}^{(1,2)}&\ldots&\ldots&\ldots&\ldots&{\bf E}^{(1,n)}  \\
{\bf E}^{(2,1)} &\ddots&\vdots&\ldots&\vdots&\vdots&\vdots \\
\vdots&\ldots&{\bf V}^{(j)}&\ldots&{\bf E}^{(j,k)}&\ldots&\vdots \\
\vdots&\vdots&\vdots&\ddots&\vdots&\vdots&\vdots \\
\vdots&\ldots&{\bf E}^{(k,j)} &\ldots&{\bf V}^{(k)}&\ldots&\vdots\\
\vdots&\vdots&\vdots&\ldots&\vdots&\ddots&\vdots \\ 
{\bf E}^{(n,1)}&\ldots&\ldots&\ldots&\ldots&\ldots&{\bf V}^{(n)}
            \end{pmatrix}, 
\end{equation}
\end{widetext}
where ${\bf V}^{(i)}$ stands for the covariance matrix of the (initial) reduced
$i^{\underline{\text{th}}}-$mode state $\hat \varrho^{(i)}\equiv\Tr_{\{l\}}(\hat \rho_G)$ 
(with $l=1,\ldots,n$ such that $l\neq i$), and ${\bf E}^{(j,k)}$ denotes the intermodal correlation 
matrix between the modes $j$ and $k$; 
note that since $\bf V$ is symmetric, ${\bf E}^{(k,j)} = ({\bf E}^{(j,k)})^\top$.  
With this notation, 
the covariance matrix of the two ($j$ and $k$) modes ---corresponding to the reduced state 
$\hat \varrho^{(jk)}=\Tr_{\{l\}}(\hat \rho_G)$, with $l=1,\ldots,n$ such that $l\neq j,k$--- reads 
\begin{equation}
\label{V(jk)}
{\bf V}^{(j,k)} = \left(\begin{array}{cc}
                {\bf V}^{(j)} & {\bf E}^{(j,k)} \\
                {\bf E}^{(k,j)} & {\bf V}^{(k)}
                \end{array} \right).
\end{equation}
\par
Our method for determining the matrix ${\bf V}$ is based on determining first the elements of the matrices ${\bf V}^{(i)}$, and then the elements of the matrices ${\bf E}^{(j,k)}$, as follows.
\subsection{Determination of the reduced single-mode covariance matrices
of the initial state $\hat \rho_G$}
\label{Va}
In order to relate the total phase $\phi_{\sf S}[\hat \rho_G]$ with single-mode covariance matrices of the initial state $\hat \rho_G$, we apply local metaplectic operations with respect to the modes $j$ and $k$, associated with symplectic matrices ${\sf S}$
of the form:
\bea                                                                                     \label{generic-S-block}
\mathsf S = {\sf I}_{2j-2} \oplus {\sf S}^{(j)} 
                           \oplus {\sf I}_{2k - 2j - 2} 
                           \oplus {\sf S}^{(k)} \oplus{\sf I}_{2n-2k}.
\eea
%
%
Resorting to Eqs.~(\ref{SimpJ}) and (\ref{Cayley1}) this gives 
\begin{equation}                                                                        \label{generic-CS-block}
{\bf C}_\mathsf{S}  = {\bf 0}_{2j-2} \! \oplus {\bf C}_{{\sf S}^{(j)}} \!
                                            \oplus {\bf 0}_{2k - 2j - 2}   \!
                                            \oplus {\bf C}_{{\sf S}^{(k)}} \!
                                            \oplus {\bf 0}_{2n-2k}, 
\end{equation} 
%
%
where ${\bf 0}_{j}$ is the $ j \times j$ null matrix, 
and ${\bf C}_{{\sf S}^{(j)}}=
\mathsf{J}_2^{(j)} \left(  \mathsf{I}_{2} - {\sf S}^{(j)} \right)
           \left(  \mathsf{I}_{2} + {\sf S}^{(j)} \right)^{-1}$. The block matrix ${\bf V}^{(j)}$ (equivalently ${\bf V}^{(k)}$) 
can be selected  by choosing            
${\sf S}^{(k)}={\sf I}_2$ (equivalently ${\sf S}^{(j)}={\sf I}_2$) 
in Eq.~(\ref{generic-S-block}). 
This corresponds to evolve the single mode $j$ (or $k$), 
and consequently the total phase $\phi_{\sf S}[\hat \rho_G]$ 
depends only on the reduced state $\hat \varrho^{(j)}$ (equivalently $\hat \varrho^{(k)}$). 
As long as none of the eigenvalues of ${\sf S}^{(k)}$ (equivalently ${\sf S}^{(j)}$) equals $-1$, 
the total phase results in 
\beq
\label{phi-reduced-A}
\phi_{{\sf S}^{(k)}}[\hat \varrho^{(k)}]= \frac{\pi}{2} \nu^+_{{\mathsf S}^{(k)}} - 
                  \frac{1}{2}{\rm arg}\left[
                                           \det(\tfrac{1}{2}{\mathsf I}_{2} + 
                                           i {\bf V}^{(k)}{\bf C}_{{\mathsf S}^{(k)}}) \right], 
\eeq
and similarly for the mode $j$. 
Naturally, Eq.~(\ref{phi-reduced-A}) is the single-mode version of Eq.~(\ref{TotalPha3}), 
%
%
and since ${\bf V}^{(k)}$ is a $2\times 2$ matrix,  Eq.~(\ref{phi-reduced-A}) reduces to 
\bea                                                                
&\phi_{{\sf S}^{(k)}}[\hat \varrho^{(k)}] 
=\frac{\pi}{2} \nu^+_{{\sf S}^{(k)}}-\nonumber\\
      &\frac{1}{2}{\rm arg}\left[ \tfrac{1}{4}\! - 
                                   \! \det({\bf V}^{(k)}{\bf C}_{{\sf S}^{(k)}})
                                   + \tfrac{i}{2} {\rm Tr}({\bf V}^{(k)}{\bf C}_{{\sf S}^{(k)}}) 
                             \right]\!\!.   \label{TotPha1M}                                                                                                                 
\eea  %
This expression allows us to determine the elements of the matrix 
\begin{equation}
\label{V-one-mode}
{\bf V}^{(k)} = \left(\begin{array}{cc}
                a & c \\
                c & b
                \end{array} \right),
\end{equation}
with $a,b>0$, once appropriate one-mode metaplectic evolutions are implemented, 
and the phase $\phi_{{\sf S}^{(k)}}[\hat \varrho^{(k)}]$ is known. 

In what follows we describe three strategies to do so. 
The single-mode transformations involved are: rotation (${\sf R}$), 
squeezing (${\sf Z}$), position shear (${\sf F}$), and momentum shear (${\sf M}$), corresponding to the Hamiltonians 
\bse
\label{one-mode-Hamiltonians}
\bea
\hat H_{\sf R}&=&\hbar \omega (\hat a^\dagger\hat a+1/2), 
\label{Hamil-rotation}\\
\hat H_{\sf Z_{\varphi}}&=& \frac{\hbar \omega}{2}(\hat a^{\dag 2}e^{i\varphi} + e^{-i\varphi}\hat a^{2}), 
\label{Hamil-squeezing}\\
\hat H_{\sf F} &=& -\frac{\hbar\omega}{4} (\hat a^{\dag} - \hat a)^2,\; \mbox{and}\; 
\label{Hamil-shear-position}\\
\hat H_{\sf M}&=& \frac{\hbar\omega}{4} (\hat a^{\dag} + \hat a)^2,
\label{Hamil-shear-momentum} 
\eea
\ese
respectively. All these metaplectic transformations are described in detail in the Appendix \ref{examples}, 
where the total phases acquired by the reduced (single-mode) Gaussian state 
for each evolution are shown to be (here $\tau = {\rm det} {\bf V}^{(i)} = ab-c^2\ge 1/4$, 
and $\beta = {\rm Tr} {\bf V}^{(k)} = a+b > 0 $):
\bea
                                                                         \label{TPRotA}
\!\!\!\phi_{\sf R}\!  =\! \frac{\pi}{2} \nu^+_{\mathsf R} \! - \!
                \frac{1}{2}{\rm arg}\left[ \tfrac{1}{4}\! -\! \tau
                {\tan}^2\tfrac{\theta}{2} \!+\! \frac{i}{2} \beta {\tan}\tfrac{\theta}{2} 
                                    \right]\!\!                                                                                                            
\eea
whenever $\theta=\omega t \neq \pi, 3\pi$;
\bea                                                                       \label{TPSqA}
&\phi_{{\sf Z}_{\varphi}}  = 
- \frac{1}{2}{\rm arg}\left[ \tfrac{1}{4}\! + \! 
   {\tau} \tanh^2\!\tfrac{\zeta}{2}+\right.\nonumber\\
&+ \left. i (\tfrac{a-b}{2} \cos\varphi + c \sin\varphi) \tanh\!\tfrac{\zeta}{2}                                             
                                    \right],                                                                                                             
\eea
where $\zeta = \omega t$ is the squeezing parameter;
\begin{eqnarray}                                                                         \label{TPCshA}
\phi_{\sf F}  = - \frac{1}{2}{\rm arg}\left(1 + i b s\right),                                                                                                                                           
\end{eqnarray}
where $s=\omega t \geq 0$; and
\begin{eqnarray}                                                                         \label{TPMshA}
\phi_{\sf M}  = - \frac{1}{2}{\rm arg}\left(1 + i a s\right),                                                                                                                                           
\end{eqnarray}
where $s=\omega t \geq 0$.

Under a particular metaplectic evolution, the initial single-mode Gaussian state
$\hat \varrho^{(k)}$ acquires a total phase given by either one of Eqs.~(\ref{TPRotA})-(\ref{TPMshA}). 
Such a phase depends on the evolution parameter as well as on the elements of ${\bf V}^{(k)}$. 
Therefore, a single evolution (hence knowledge of a single $\phi$) does not suffice 
to invert the equations and completely determine all the elements of ${\bf V}^{(k)}$. 
Thus, a set of evolutions over the initial state (or rather, a set of phases) is needed to determine ${\bf V}^{(k)}$. 

For each evolution an acquired $\phi$ is determined, all of which depend on the same (initial) covariance matrix. 
Ultimately, when a sufficient number of phases are known, 
this allows for the inversion of the set of equations and the determination of all the elements of ${\bf V}^{(k)}$.

\underline{\it First strategy.}
This strategy is more suitable to be used in the determination of Gaussian states within the context of CV systems corresponding to vibrational modes of trapped ions. 
It involves the application of two different rotations,  and two different 
squeezing transformations.
\par
For $0\leq \theta < \pi$  we have $\nu^+_{\sf R}=0$ (see Eq.~(\ref{nu-plus-rot})
in Appendix \ref{examples}). Then, according to Eq.~(\ref{TPRotA}),  
$\tan(-2\phi_{\sf R})=\tan({\rm arg}\left[ z\right])$ with 
$z=\tfrac{1}{4}\! -\! \tau
                {\tan}^2\tfrac{\theta}{2} \!+\! \frac{i}{2} \beta {\tan}\tfrac{\theta}{2}$ and
                $\Im(z)\ge 0$.  Therefore, $z$ could be in the first or second quadrant 
                of the complex plane. In both cases we have
\begin{eqnarray}                                                                         \label{TPRot1}
\tan (-2 \phi_{\sf R})  =\frac{\Im[z]}{\Re[z]}=  
\frac{2\beta \tan\tfrac{\theta}{2}}{1-4\tau \tan^2\tfrac{\theta}{2}}.                                          
\end{eqnarray}
Assume that 
two values of the total phase, namely
$\phi'_{\sf R} := \phi_{\sf R}(\theta')$ and 
$\phi''_{\sf R} := \phi_{\sf R}(\theta'')$
are known, corresponding to two distinct rotation angles $\theta'$ and $\theta''$, both in the interval $[\pi/2,\pi)$. Then, substituting these two values in Eq.~(\ref{TPRot1}), 
we can set up a linear system in the variables $\tau$ and $\beta$, whose solution is
\bse
\begin{eqnarray}                                                                         
&&\beta  = \frac{ \left( \cot^2\!\tfrac{\theta}{2}' - \cot^2\!\tfrac{\theta}{2}'' \right) \tan(2 \phi_{\mathsf R}')  \tan(2 \phi_{\mathsf R}'') }
                { 2 \cot\!\tfrac{\theta}{2}'' \tan(2 \phi_{\mathsf R}')-2\cot\!\tfrac{\theta}{2}' \tan(2 \phi_{\mathsf R}'')}, \label{TPRot2} \\                
&&\tau  = \frac{ \cot\!\tfrac{\theta}{2}'\tan(2 \phi_{\mathsf R}')-\cot\!\tfrac{\theta}{2}''\tan(2 \phi_{\mathsf R}'')}
                { 4 \tan\!\tfrac{\theta}{2}' \tan(2 \phi_{\mathsf R}')- 4\tan\!\tfrac{\theta}{2}'' \tan(2 \phi_{\mathsf R}'')} \label{TPRot22}.
\end{eqnarray}
\ese
\par
On the other hand, the real part of the complex number 
$z^\prime$
in the argument function in Eq.~(\ref{TPSqA}) 
is always positive. So, $z^\prime$ could be in the first or fourth quadrant of the complex plane. In both cases we have
\bea
&\tan (-2\phi_{{\sf Z}_{\varphi}}) = \frac{\Im[z^\prime]}{\Re[z^\prime]}=\nonumber\\
&=\frac{2(a-b)\cos\varphi + 4c \sin\varphi }
{1 + 4 \tau \tanh^2\tfrac{\zeta}{2}} \tanh\!\tfrac{\zeta}{2}.
\label{TgphiZ} 
\eea
%
%
Once the value of $\tau$ has been obtained from Eq.~(\ref{TPRot22}), 
it only remains to perform a squeezing transformation 
(with squeezing parameter $\zeta$), 
and determine the value of the phase $\phi_{{\sf Z}_{\varphi}}$ 
to obtain $c$ from Eq.~(\ref{TgphiZ}) with $\varphi = \pi/2$:
\begin{equation}
c = - \frac{ 1 + 4 \tau \tanh^2\tfrac{\zeta}{2} }{4 \tanh\!\tfrac{\zeta}{2}} \tan (2\phi_{{\sf Z}_{\frac{\pi}{2}}}).
\end{equation}
Finally, applying a second squeezing transformation with the same squeezing parameter 
$\zeta$, but now with $\varphi=0$, we get from Eq.~(\ref{TgphiZ}) the following value 
for $\gamma:=b-a$ 
\beq
\gamma=
\frac{1 + 4 \tau \tanh^2\tfrac{\zeta}{2}}
{2\tanh\!\tfrac{\zeta}{2}}\tan (2\phi_{{\sf Z}_{0}}).
\eeq
Then, as we have $a+b=\beta$ and $b-a=\gamma$, we can calculate $a=(\beta-\gamma)/2$
and $b=(\beta+\gamma)/2$, and in this way completely determine the covariance matrix (\ref{V-one-mode}).
\par
\noindent
\par
\underline{\it Second strategy.}
This strategy is more suitable to be used in the determination of Gaussian states within the context of CV systems corresponding to 
quadrature-modes of the quantized electromagnetic field. 
It relies on three different rotations and a squeezing transformation, 
but 
here only the total phases corresponding to the rotations have to be determined, as we shall see.

First, the trace $\beta$ and the determinant $\tau$ of the matrix (\ref{V-one-mode}) 
are calculated by performing two different rotations, exactly as we did in Eqs.~(\ref{TPRot2}) and (\ref{TPRot22}). 
Then, a unitary evolution corresponding to a squeezing transformation is performed over the state $\hat \rho_{G}$ 
with Hamiltonian (\ref{Hamil-squeezing}) setting $\varphi=0$, so the evolved state will be 
$\hat \rho'_{G} = \hat M_{{\sf Z}_{0}} \hat \rho_{G} \hat M^\dagger_{{\sf Z}_{0}}$.
The corresponding symplectic transformation is given by (\ref{squeezing}), 
leading to a covariance matrix of the evolved state equal to ${\bf V}' = {\sf Z}_{0}{\bf V}{\sf Z}_{0}^\top$. 

Now, a third rotation of an angle $\theta''' \in [\pi/2,\pi)$ is performed to obtain the phase using (\ref{TPRot1}) 
for the new (squeezed) state $\hat \rho'_{G}$. 
Thus, defining $\phi'''_{\sf R} := \phi_{\sf R}(\theta''')$, one gets 
\begin{eqnarray}                                                                         \label{TPR2}
\tan (2 \phi'''_{\sf R})  =  \frac{2\beta' \tan\tfrac{\theta}{2}}{4\tau \tan^2\tfrac{\theta}{2} - 1},                                          
\end{eqnarray}
where we have used ${\rm det} {\bf V}' = {\rm det} {\bf V} = \tau$, and
$\beta' := {\rm Tr} {\bf V}' = \beta \cosh(2 \zeta) - 2 c \sinh(2 \zeta)$.
Solving Eq.~(\ref{TPR2}) for $c$, one finds 
\begin{equation}
c =  \frac{1 - 4\tau \tan^2\tfrac{\theta}{2}}{ 4 \tan\tfrac{\theta}{2} \sinh(2\zeta)} \tan (2 \phi'''_{\sf R}) 
     + \frac{\beta}{2} {\rm cotanh}(2\zeta). 
\end{equation}
With this, and $\beta$ and $\tau$ given by Eqs.~(\ref{TPRot2}) and (\ref{TPRot22}), respectively, the system of equations $\beta = a + b$ and $\tau = ab -c^2$ can be solved to get $a = \sqrt{\beta -\tau -c^2}$ and $b = \beta - \sqrt{\beta -\tau -c^2}$. 
It is worth noting that, in this strategy, it is necessary to determine only phases associated with rotations. 
The evolution phase corresponding to the intermediate  application of a squeezing transformation does not need to be determined.

\underline{\it Third strategy.} 
Although the first and second strategies can be implemented in the determination of Gaussian states within the context of CV systems corresponding to spatial transverse degrees of freedom of single photons,
this third strategy would be experimentally less demanding in this particular system.
It involves the implementation of two squeezing plus a coordinate or momentum shear transformation. 

Let us assume that two values of the total phase, namely
$\phi^\prime_{{\sf Z}} := \phi_{{\sf Z}_{\frac{\pi}{2}}}(\zeta^\prime)$ and 
$\phi^{\prime\prime}_{\sf Z} := \phi_{{\sf Z}_{\frac{\pi}{2}}}(\zeta^{\prime\prime})$ are known,
corresponding to squeezing transformations for two distinct values of the squeezing parameter $\zeta$, and $\varphi = \pi/2$.
Then, from Eq.~(\ref{TgphiZ}) we can set up a linear system of equations with unknown 
variables $c$ and $\tau$, whose solution is
\bse
\bea
& \, \nonumber\\
&\!\! \!\! \!\! \!\! \! \! c=\frac{\frac{1}{4}\!\left[ \tanh^2\!\tfrac{\zeta''}{2}\!-\! 
                                                         \tanh^2\!\tfrac{\zeta'}{2}\right]\!\!
                                    \tan(2\phi^{\prime}_{\sf Z })\!\tan(2\phi^{\prime\prime}_{\sf Z })}
{\tanh\!\tfrac{\zeta''}{2}\tanh\!\tfrac{\zeta'}{2}\!\!
\left[\!\tanh\!\tfrac{\zeta''}{2}\!\tan(2\phi^{\prime\prime}_{\sf Z }) 
      \!-\!\tanh\!\tfrac{\zeta'}{2}\!\tan(2\phi'_{\sf Z })\!\right]},\\
& \nonumber\\
&\!\! \!\! \!\! \!\! \! \!\tau= \frac{\frac{1}{4}\tanh\!\tfrac{\zeta'}{2}\tan(2\phi^{\prime\prime}_{\sf Z }) - 
               \frac{1}{4}\tanh\!\tfrac{\zeta''}{2}\tan(2\phi'_{\sf Z })}
{\tanh\!\tfrac{\zeta''}{2} \! \tanh\!\tfrac{\zeta'}{2}\!\!
\left[\!\tanh\!\tfrac{\zeta'}{2}\!\tan(2\phi'_{\sf Z })\!-\!
\tanh\!\tfrac{\zeta''}{2}\!\tan(2\phi^{\prime\prime}_{\sf Z })\!\right]}.
\eea
\ese
\par
In order to determine the matrix elements $a$ and $b$ 
we need to perform either a position or a momentum shear, and use Eqs.~(\ref{TPCshA}) or (\ref{TPMshA}). 
Thus, for example, if we perform a position shear so that $\phi_{\sf F}$ is known, Eq.~(\ref{TPCshA}) leads to 
\begin{equation}
\label{valor-b}
b = -\frac{1}{s}\tan (2\phi_{\sf F}).
\end{equation}
Then, once $b$, $c$, and $\tau$ are known, it is straightforward to determine $a$ according to $a=(\tau+c^2)/b$. 
Alternatively, we can perform a momentum shear transformation, determine the total phase $\phi_{\sf M}$, and resort to 
Eq.~(\ref{TPMshA}) to get 
\begin{eqnarray}                                                                         \label{value-a}
a = -\frac{1}{s}\tan (2\phi_{\sf M}).                                                                                                                                        
\end{eqnarray}
Once $a$, $c$, and $\tau$ are known, we can obtain the matrix element $b$ according to $b=(\tau+c^2)/a$.

\subsection{Determination of the two-mode intermodal correlation matrices
of the initial state $\hat \rho_G$}
\label{Vb}
The preciding section provided a method for determining the one-mode covariance matrices 
${\bf V}^{(i)}$ with $i=j,k$. Here we will assume that these two matrices are already known, 
and develop a strategy for determining the elements of a generic two-mode correlation matrix
\beq \label{coormatrix}
{\bf E}^{(j,k)}  = \begin{pmatrix} 
      v & w \\
      y & z \\
   \end{pmatrix}.
\eeq

The method exhibits the same spirit as that for determining ${\bf V}^{(k)}$ in the sense that it resorts to the implementation of appropriate metaplectic evolutions to extract information regarding the matrix elements of ${\bf E}^{(j,k)}$. However, it differs from the strategies of Section \ref{Va} in that here, before the determination of the evolution phases, an extra  two mode and single mode rotations must be implemented.
\par
We first apply a (non-local)
two-mode rotation, corresponding to the Hamiltonian 
$\hat H^{(j,k)}=(\hbar\omega/2)(\hat a_{k}^\dagger\hat a_j+\hat a_k\hat a_j^\dagger)$ 
(when the continuous variable system refers to the 
quantized electromagnetic fields, $\hat H^{(j,k)}$ represents the beam splitter 
evolution over modes $j$ and $k$).
If the rotation is performed by an angle $\theta=\omega t=\pi/2$, 
the symplectic matrix associated with the $n-$mode transformation reads 
\begin{equation} \label{generic-S-block-2}                                                                        
{\sf S} = {\sf I}_{2j-2} \oplus {\sf S}^{(j,k)} \oplus {\sf I}_{2n-2k},
\end{equation} 
with the $(2k-2j+2) \times (2k-2j+2)$ matrix ${\sf S}^{(j,k)}$ given by
\beq \label{Sjk-rot}
{\sf S}^{(j,k)} = \frac{1}{\sqrt{2}}
\begin{pmatrix}
{\sf I}_2  & {\bf 0}_2 & \ldots & {\bf 0}_2 & {\sf J}_2  \\
{\bf 0}_2  & {\sf I}_2 & \ldots & {\bf 0}_2 & {\bf 0}_2 \\
\vdots     & \vdots    & \ddots & \vdots    & \vdots \\
{\bf 0}_2  & {\bf 0}_2 & \ldots & {\sf I}_2 & {\bf 0}_2  \\
{\sf J}_2  & {\bf 0}_2 & \ldots & {\bf 0}_2 & {\sf I}_2
\end{pmatrix}.
\eeq
This matrix corresponds to the unitary metaplectic evolution 
$\hat M_{{\sf S}^{(j,k)}}$ corresponding to a two mode rotation by angle $\omega t=\pi/2$. 
The total ($n-$mode) metaplectic evolution associated with ${\sf S}$ 
in Eq.~(\ref{generic-S-block-2}) is thus 
$\hat M_{\sf S} = \hat{\sf 1}_{j-1} \otimes \hat M_{{\sf S}^{(j,k)}} \otimes \hat{\sf 1}_{n-k}$,
where $\hat{\sf 1}_{j-1}$ is the identity operator acting on the first $j-1$ modes. 
and the evolved state is $\hat \rho'_G = \hat M_{{\sf S}} \hat \rho_G\hat M_{{\sf S}}^\dagger$. 

Let ${\bf V^\prime}^{(j,k)}$ denote the covariance matrix of the reduced two-mode evolved state 
$\hat \varrho^{\prime(jk)}=\Tr_{\{l\}}(\hat \rho'_G)$, 
with $l=1,\ldots,n$ such that $l\neq j,k$. 
Such a matrix is related to the original (non-evolved) covariance matrix 
${\bf V}^{(j,k)}$ in (\ref{V(jk)}) 
according to 
\begin{equation}
{\bf V}^{'(j,k)} = \frac{1}{2} 
\begin{pmatrix}
\mathsf I_2 & {\sf J}_2  \\
{\sf J}_2 & \mathsf I_2  
\end{pmatrix}
{\bf V}^{(j,k)}
\begin{pmatrix}
\mathsf I_2 & {\sf J}_2  \\
{\sf J}_2 & \mathsf I_2  
\end{pmatrix}^\top . 
\end{equation} 
The diagonal blocks of the above matrix are the 
single-mode covariance matrices given by 
\bse
\label{redVprime-sys}
\bea
&\!\!\!\!\!\!\!\!{\bf V^\prime}^{(j)}& = {\bf V}^{(j)} \! + \! {\sf J}_2 {\bf E}^{(k,j)}
                                         -\!{\bf E}^{(j,k)}\!{\sf J}_2 \! - 
                                         \! {\sf J}_2 \! {\bf V}^{(k)} \! {\sf J}_2,                      \\
&\!\!\!\!\!\!\!\!{\bf V^\prime}^{(k)}& = {\bf V}^{(k)} \! - \! {\sf J}_2 \! {\bf V}^{(j)} \! {\sf J}_2 \!
                                         - \! {\bf E}^{(k,j)} \! {\sf J}_2 \!
                                         + \! {\sf J}_2 {\bf E}^{(j,k)},
\eea
\ese
corresponding to the reduced single-mode states
$\hat \varrho^{(i)}\equiv\Tr_{\{l\}}(\hat \rho'_G)$ (with $l=1,\ldots,n$ such that $l\neq i=j,k$).
Solving Eqs.~(\ref{redVprime-sys}) for ${\bf E}^{(j,k)}$, 
we obtain 
\beq
\label{E1}
2{\sf J}_2 {\bf E}^{(k,j)} {\sf J}_2 + 2{\bf E}^{(j,k)} = {\bf V^\prime}^{(j)}{\sf J}_2-
{\sf J}_2{\bf V^\prime}^{(k)}.
\eeq
Once the matrices ${\bf V^\prime}^{(j)}$ and ${\bf V^\prime}^{(k)}$ 
are determined using some strategy, as explained above, 
Eq.~(\ref{E1}) becomes a linear system for the matrix elements of ${\bf E}^{(j,k)}$ in (\ref{coormatrix}). 
From this system, one is able to obtain
\bse
\label{result-yw}
\bea
w &=&\frac{1}{4}[{\bf V^\prime}^{(j)}{\sf J}_2- {\sf J}_2{\bf V^\prime}^{(k)}]_{1,2} \, ,\\
y&=&\frac{1}{4}[{\bf V^\prime}^{(j)}{\sf J}_2 - {\sf J}_2{\bf V^\prime}^{(k)}]_{2,1}\, .
\eea
\ese
In order to determine
$v$ and $z$, 
we perform an additional (single-mode local) operation over the mode $j$ of the evolved state $\hat \rho'_G$, 
with the symplectic matrix ${\sf J}_2^{(j)}$, so that
\begin{equation}
{\sf S}' = {\sf I}_{2j-2} \oplus \left[ {\sf S}^{(j,k)} ({\sf J}_2 \oplus {\sf I}_{2k - 2j}) \right] 
                          \oplus {\sf I}_{2n-2k}
\end{equation} 
%
with ${\sf S}^{(j,k)}$ given in Eq.~(\ref{Sjk-rot}).
The single-mode operation 
$\hat{\sf 1}_{j-1} \otimes \hat M_{{\sf J}^{(j)}_2} \otimes \hat{\sf 1}_{n-j}$
is implemented via a rotation with the angle $\theta = \pi/2$ 
(see Appendix {\rm Aa}), and the evolved state is 
$\hat \rho''_G = \hat M_{\sf S''}\hat \rho'_G\hat M^\dagger_{{\sf S''}}$, 
where now 
$\hat M_{{\sf S''}} = \hat{\sf 1}_{j-1} \otimes \hat M_{{\sf J}^{(j)}_2} \otimes \hat{\sf 1}_{n-j}$.
Denoting by ${\bf V^{\prime\prime}}^{(i)}$ the covariance matrix of mode $i$ after the evolution, we proceed as we did to arrive at Eqs.~(\ref{redVprime-sys}) and get 
\bse
\label{redVprime-sys-2}
\bea
&\!\!\!\!\!\!\!\!{\bf V^{\prime\prime}}^{(j)}& =  {\sf J}_2 \! \left[ {\bf V}^{(j)} \!+\! 
                                                                      {\bf V}^{(k)} \!+\! 
                                                                      {\bf E}^{(j,k)} \!+\!
                                                                      {\bf E}^{(k,j)} \right]\! 
                                                  {\sf J}_2^\top,  \\
&\!\!\!\!\!\!\!\!{\bf V^{\prime\prime}}^{(k)}& = {\bf V}^{(j)} -{\bf E}^{(k,j)}-{\bf E}^{(j,k)}+{\bf V}^{(k)}.
\eea
\ese
Solving Eqs.~(\ref{redVprime-sys-2}) for ${\bf E}^{(j,k)}$, 
we are led to the linear system
\beq
-2({\bf E}^{(k,j)} + {\bf E}^{(j,k)})=
{\sf J}_2{\bf V^{\prime\prime}}^{(j)}{\sf J}_2+
{\bf V^{\prime\prime}}^{(k)},
\eeq
from which we obtain the matrix elements
\bse
\label{result-xz}
\bea
v&=&-\frac{1}{4}[{\sf J}_2{\bf V^{\prime\prime}}^{(j)}{\sf J}_2+
{\bf V^{\prime\prime}}^{(k)}]_{1,1}\\
z&=&-\frac{1}{4}[{\sf J}_2{\bf V^{\prime\prime}}^{(j)}{\sf J}_2+
{\bf V^{\prime\prime}}^{(k)}]_{2,2}. 
\eea
\ese
Therefore the elements $w,y,v$, and $z$ of the matrix ${\bf E}^{(j,k)}$ 
are written in terms of single-mode CMs, which can be determined using some strategy developed in Sec.\ref{Va}.
%
\par
Gathering results, we have provided a method that allows us to determine any (all) two-mode covariance matrices (\ref{V(jk)}) with due implementation of one- and two-mode metaplectic evolutions. By applying the method repeatedly (varying $j$ and $k$), the complete covariance matrix (\ref{V-block}) of an arbitrary Gaussian state can be determined. 
\par
Finally, it is important to notice that the strategies developed here, involving the unitary operations as simply as possible, are suitable for several paradigmatic CV systems. In general, the same procedure can be applied with 
any combination of metaplectic evolutions (or equivalently, quadratic Hamiltonians), 
and the same results are obtained with alternative designed strategies, as long as the new set of evolutions allows one to extract the covariance matrix
${\bf V}$ from Eqs.~(\ref{total-phase-total-system})

\subsection{Determination of Entanglement in pure Gaussian states}
\label{VI}
\par
Though all the informational properties of a GS are contained in its covariance matrix, 
in certain cases partial information of the full CM suffices to extract information regarding the entanglement between the modes.
For example, in the case of $n$-mode pure Gaussian states, $\hat \rho_G^{{\rm pure}}$,
the amount of entanglement 
in an arbitrary bipartition $A|B$ with $n_A\times n_B$ modes (such that $n_A + n_B=n$),
can be computed resorting only to the covariance matrix of any of the 
reduced Gaussian states, namely $\hat\varrho_A$ or $\hat\varrho_B$
\cite{Adesso2014}. In this regard, our method, like other methods in general,
allows us to determine such reduced covariance matrices. 
However, it has the advantage that it provides an experimentally friendly way to determine the purity of the reduced single-mode states ---whence allows to measure the entanglement in bipartitions
 having $1\times(n-1)$ modes--- without the need to determine the full
reduced covariance matrix.
\par  
The purity of the reduced $i^{\underline{\text{th}}}-$mode Gaussian state $\hat {\varrho}^{(i)}$ is given by 
\beq
\Tr(\hat {\varrho}^{(i)})^2 = \frac{\hbar}{2\sqrt{ {\rm det} {\bf V}^{(i)} }}=\frac{\hbar}{2\sqrt{\tau}},
\eeq
and can be determined, according to Eq.~(\ref{TPRot22}), 
from the knowledge of the total phases associated with only two local rotations, which determine the value of $\tau$. Once $\tau$ is known, 
the amount of entanglement ${\mathcal E}$ between the $i^{\underline{\text{th}}}$ and the remaining $n-1$ modes can be computed using the pure-state Rényi entropy of entanglement \cite{Adesso2014,adesso2012}
\beq
{\mathcal S}_{\alpha}(\hat {\varrho}^{(i)})=
\frac{\ln(\Tr(\hat {\varrho}^{(i)})^\alpha)}{(1-\alpha)}
\eeq
with $\alpha=2$, which gives
\bea
{\mathcal E}={\mathcal S}_2(\hat {\varrho}^{(i)}) 
&=&(1/2)\ln(\tau)-\ln(\hbar/2).
\eea

\par
 
\section{Measurement protocol of the total phase}
\label{VII}
In this section we describe our protocol to measure both the real and the imaginary parts 
of $\textrm{Tr}(\hat \rho\, \hat U)$, in order to compute $\phi$ resorting to Eq.~(\ref{Totpha}).
The main idea is to entangle the $n-$mode system in an arbitrary state $\hat \rho$ with a 
qubit ancilla, using the conditional evolution 
\beq
\label{Ucond}
\hat U^{(c)}\equiv \exp[-i\frac{t}{2\hbar}(\hat 1+\hat{\sigma}_3) \otimes \hat H],
\eeq 
with $\hat H$ an arbitrary Hamiltonian 
acting on the $n-$mode system.
We use  $\ket{j,\pm}$ to denote the eigenstates of the Pauli operators $\hat{\sigma}_j$ ($j=1,2,3$), so that $\hat{\sigma}_j\ket{j,\pm}=\pm\ket{j,\pm}$.  
\par 
Initially, the $n-$mode system and the ancilla are assumed to be in the separable state $\dyad{1,+}{1,+}\otimes\hat\rho$, with $\ket{1,+}=(1/\sqrt{2})(\ket{3,+}+\ket{3,-})$. 
The reduced state of the qubit ancilla after the evolution of the complete ($n-$mode plus ancilla) system is thus given by 
\bea
\label{total-state}
\hat \rho_q &=& \Tr_{n}\left[\hat U^{(c)}\dyad{1,+}{1,+}\otimes\hat\rho\;\hat U^{(c)\dagger}\right] \nonumber\\
            &=& \frac{1}{2} \left[\dyad{3,+}{3,+} + \Tr_{n}(\hat \rho \hat U) \dyad{3,+}{3,-}  \right.\nonumber\\
            & & + \left. {\rm Tr}_n ( \hat \rho \,\hat U^\dagger) \dyad{3,-}{3,+} + \dyad{3,-}{3,-} \right],
\eea
where $\Tr_{n}(\ldots)$ denotes the trace over the $n-$mode system, and $\hat U= \exp (-i\frac{t}{\hbar}\hat H)$ is a unitary evolution acting only on the $n-$ mode system.

Then, a $\pi/2$ rotation $\hat U_{\pi/2}(\vartheta)$ around an axis in the equator of the Bloch sphere that makes an angle $\vartheta$ with the $x-$axis is performed on the qubit ancilla.
From the probability measurement of the qubit's populations, 
$P_{\pm}(\vartheta):=\Tr_q[\dyad{3,\pm}{3,\pm}\hat U_{\pi/2}(\vartheta)\hat \rho_q\hat U^\dagger_{\pi/2}(\vartheta)]$ we get
\beq
\label{Pmenos-Pmas}
P_-(\vartheta)-P_+(\vartheta)=\Im\left[e^{i\vartheta}\Tr(\hat \rho\hat U)\right].
\eeq
Thus, by choosing qubit rotations with $\vartheta=0$ and $\vartheta=\pi/2$ we obtain 
the imaginary and real part, respectively, of ${\rm Tr}(\hat \rho\, \hat U)$, and hence the total phase $\phi=\arg[{\rm Tr}(\hat \rho\,\hat U)]$ can be determined. 
\par
\subsection{The total phase acquired by an evolved reduced state in a $n-$mode system}
\label{VIIa}
\par
Let us now assume that $\hat H$ in Eq.~(\ref{Ucond}) has the form $\hat H= \hat H_A\otimes \hat{1}_{B}$, where $\hat H_A$ is a Hamiltonian acting on subsystem
$A$ consisting of $m-$modes, and $\hat{1}_{B}$ is the $(n-m)\times (n-m)$ identity operator acting on subsystem $B$. In this case $\hat U$ reduces to $\hat U= \exp(-i\frac{t}{\hbar}\hat H_A )$, and we should write $\textrm{Tr}(\hat \rho\,\hat U)=\textrm{Tr}_{A}(\hat\varrho \,\hat U)$ in Eqs.~(\ref{Pmenos-Pmas}), with $\hat\varrho$ the reduced ($m-$mode) state $\hat\varrho=\textrm{Tr}_{B}\hat \rho$.

In order to measure the phase acquired by the arbitrary 
reduced state $\hat\varrho$ we need to entangle
only the $m-$modes of interest with the qubit ancilla through the conditional evolution 
(\ref{Ucond}), with $\hat H = \hat H_A\otimes \hat{1}_{B}$. 
This is the strategy required to measure the total phases that allow us to determine
the covariance matrix of an arbitrary Gaussian state $\hat\varrho_G=\textrm{Tr}_{B}\hat \rho_G$, 
considering only (as has been shown in Section \ref{V}) metaplectic evolutions of the form in (\ref{Metap}), 
such that 
$\hat H = \hat H_A = \omega(\hat x_{\!A}^\top{\bf H}\hat x_{\!A})/2$
with $x_{\!A}^\top := (q_1,p_1,...,q_{m},p_{m})$. 
In fact, it is worth noting that,
according to the method described in Section \ref{V}, we only need to implement conditional 
evolutions over one mode, {\it viz.}, 
%
\begin{equation}\label{one-mode-cond-evol}
\!\!\hat U^{(c)}  =  \exp\! 
\left[
      {-\frac{i t}{2\hbar}(\hat 1+\hat{\sigma}_3) \! \otimes \! \hat{\sf 1}_{j-1} 
                                                  \! \otimes \! \hat H^{(j)} \!
                                                  \! \otimes \! \hat{\sf 1}_{n-j}}  \right]\!,
\end{equation}
where $ \hat H^{(j)}$ is one of the Hamiltonians in
Eqs.~(\ref{one-mode-Hamiltonians}). 
This is so because the two-mode rotation and the additional single-mode operation, 
described in Section \ref{Vb} and needed to determine the $2\times2$ intermodal 
correlation matrix ${\bf E}^{(i,k)}$, 
do not need to be applied conditionally to the state of the qubit ancilla.

\section{One-mode conditional metaplectic evolutions in several CV systems.}
\label{VIII}

The feasibility of our method for determining Gaussian states depends on the possibility of 
implementing one-mode conditional evolutions such as that in Eq.~(\ref{one-mode-cond-evol}),
with $ \hat H^{(j)}$ one of the Hamiltonians in
Eqs.~(\ref{one-mode-Hamiltonians}).
Here we describe how these conditional evolutions can be implemented in the context of 
three CV systems: 
{\it (i)} the transverse spatial degree of freedom of single photons,
{\it (ii)} the vibrational modes in trapped ions, and 
{\it (iii)} the quadrature modes of the quantized electromagnetic field.
\par 
\subsection{Transverse spatial degrees of freedom of single photons}
We consider first the implementation of single mode conditional metaplectic 
evolutions in the CV system corresponding to the transverse spatial degrees of freedom (TSDF) 
of single photons propagating in the paraxial approximation. This is the CV systems of twin 
photons generated in spontaneous parametric down conversion (SPDC) \cite{Walborn-report2010,tasca11,tasca09a}. 
Highly entangled Gaussian states can be generated with twin photons 
since, to a good approximation, generalization of 
two mode squeezed states  can be performed \cite{abouraddy07,tasca09a}.
\par 
In order to determine Gaussian states in the TSDF of single photons, the less demanding 
experimental strategy is to implement the one-mode metaplectic operations 
described in the {\it third strategy} in Section \ref{Va}, involving squeezing and 
position shear transformations.
We identify the qubit ancilla with the polarization degrees of freedom of the single photon.
As customary, we associate the horizontally ($x-$direction) and vertically ($y-$direction) polarized linear states with $\ket{H}:=\ket{3,+}$ and $\ket{V}:=\ket{3,-}$. The linearly polarized states rotated $45^\circ$ in the counter-clock wise direction with respect to $x$ and $y$ are identified, respectively, with $\ket{+45^\circ}:=\ket{1,+}$ and $\ket{-45^\circ}:=\ket{1,-}$. Finally, the states $\ket{R}:=\ket{2,+}$ and $\ket{L}:=\ket{2,-}$ are put into correspondence with the right- and left- circularly polarized states. With these identifications, measuring the polarization is equivalent to measure the Pauli observables $\hat \sigma_j$ corresponding to the qubit-ancilla polarization degrees of freedom. 
The conditional evolutions, entangling the polarization and the transverse spatial 
degrees of freedom, are implemented using a spatial light modulator (SLM) that imprints a phase only on the horizontal polarization component (transverse spatial $x-$direction).
It is worth noting that the qubit rotation that leads to Eq.~(\ref{Pmenos-Pmas})
corresponds, in this context, to mapping one orientation of linear polarization to another
or to a circular one.
\begin{figure}[tb!]
\centering
\includegraphics[width=8.0cm]{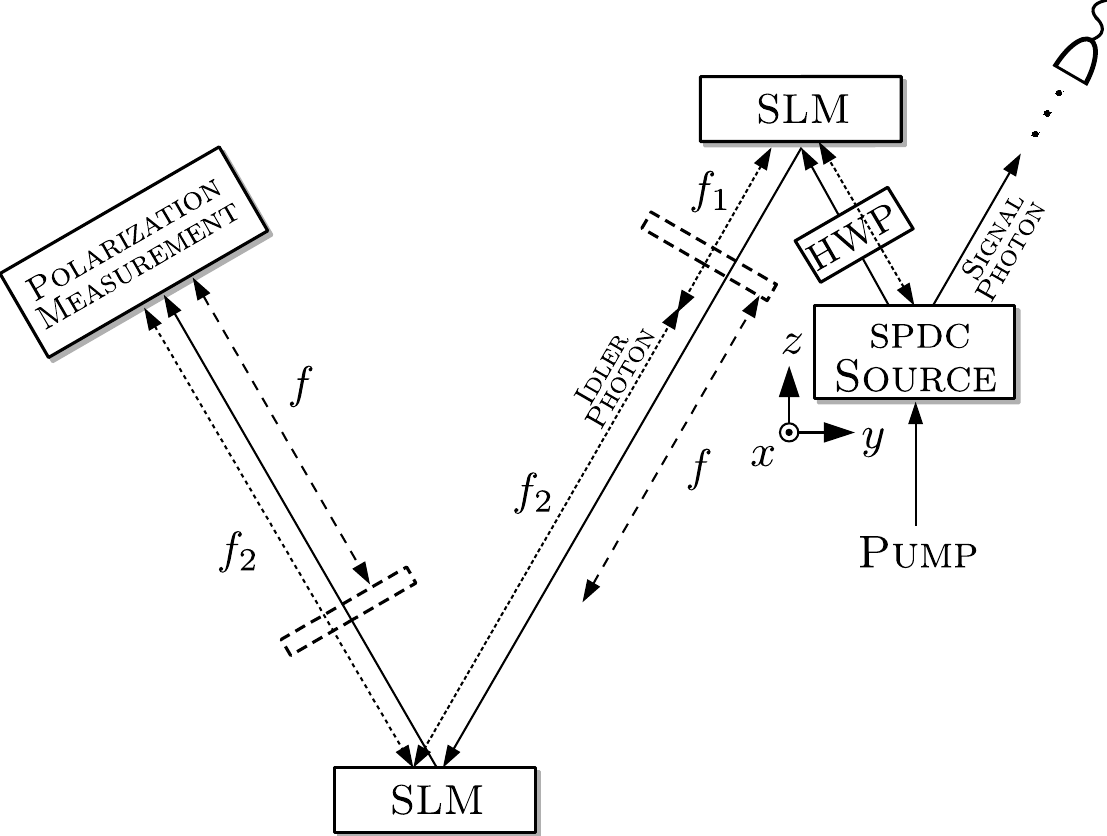}
\caption{Experimental setup to implement the conditional squeezing transformation on
the transverse spatial degree of freedom of single photons, for example generated 
with the SPDC process. 
Here, the conditional squeezing transformation is implemented on the idler photon.
On the signal photon it has to be implemented in an optical image system in both transverse degrees of freedom (not-shown) before the coincidence measurement of the twin photons.
The dashed rectangles correspond to cylindrical lenses of focal length $f$, acting 
in the transverse $y$ degree of freedom. 
The
acronym HWP denotes the half wave plate, and SLM the 
spatial light modulator. See the main text for details. }
\label{fig1}
\end{figure}

In Fig. \ref{fig1} we show the experimental setup for the implementation of 
the conditional evolution corresponding to squeezing transformation.
The half wave plate (HWP) rotates $45^\circ$ the initial polarization originally in the $x-$direction. 
The first and second SLMs, with focal lengths $f_1$ and $f_2$ respectively, 
implement thin lenses in the $x-$direction, whereas they act only as mirrors in the $y-$direction. 
Each of these lenses implements thus a Fourier transform in the $x$ spatial degree of freedom 
\cite{ozaktas01} ($\pi/2-$rotation in Lohmann's type $I$ optical configuration \cite{Lohmann1995}), 
whereas the combination produces a squeezing transformation with squeezing parameter $\zeta=f_2/f_1$ \cite{tasca11}.
The dashed rectangles represent cylindrical lenses with focal length $f$ that implement 
an optical image system in the $y$ degree of freedom, yet do not affect 
the evolution in the $x$ degree of freedom.
Notice that the total distance of propagation of the single photon until it enters 
the polarization measurement optical circuit is $4f=2f_2+2f_1$.

\par
\begin{figure}[bht]
\centering
\includegraphics[width=8.0cm]{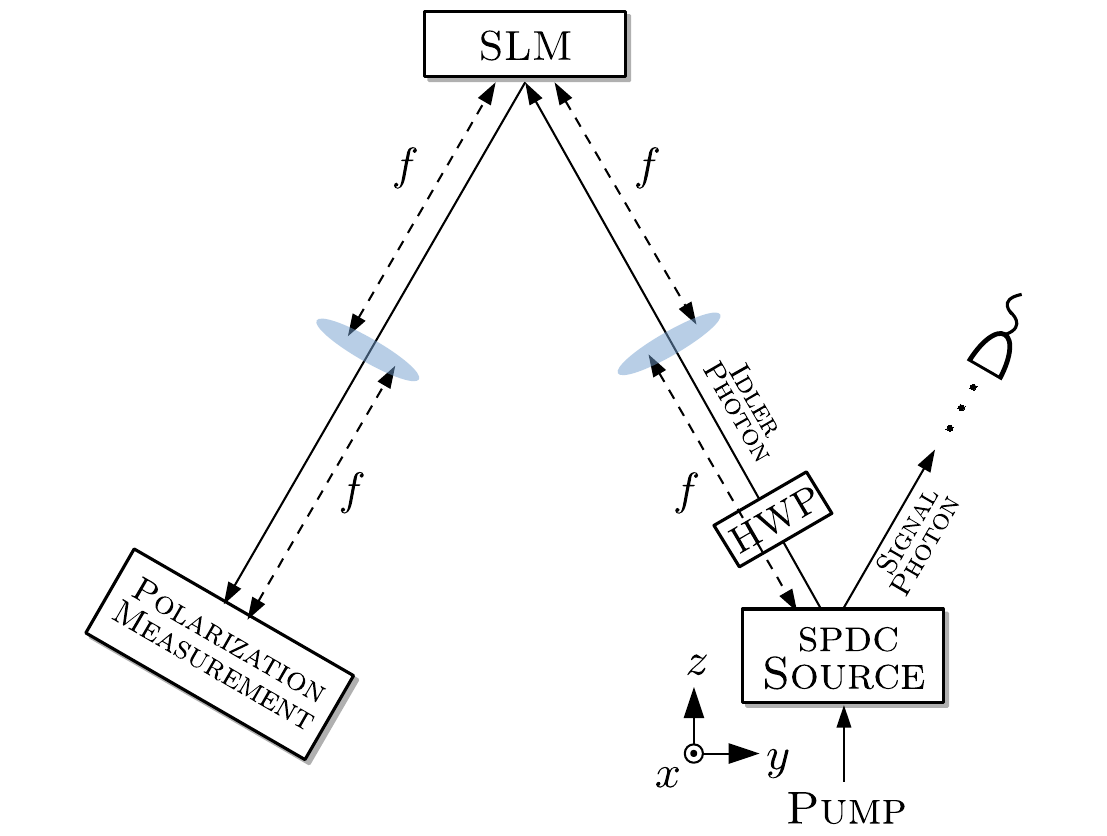}
\caption{Experimental setup to implement the conditional position shear transformation on
the transverse spatial degree of freedom of single photons, for example generated 
with the SPDC process. 
Here, the conditional position shear transformation is implemented on the idler photon.
On the signal photon it has to be implemented in an optical image system in both transverse degrees of freedom (not shown) 
before the coincidence measurement of the twin photons.
The shadded ovals correspond to spherical lenses with focal length $f$. See the main text for details.}
\label{fig2}
\end{figure}

 
In Fig. \ref{fig2} we show the experimental setup that implements the conditional 
evolution corresponding to position shear transformation.
The evolution associated with a shear in position over the single photon
is  $\braket{p}{\Phi_G^\prime}=
e^{-i\frac{\omega t}{2} p^2}\braket{p}{\Phi_G}$,
where 
$\braket{p}{\Phi_G}$ stands for the wave function in the transverse momentum representation 
at the source plane $z=0$. Thus, for a shear in position we have to map the momentum 
wave function  $\braket{p}{\Phi_G}$ at $z=0$ to the position wave 
function $\braket{x^\prime}{\Phi_G}$ in the far field, that is, when $x^\prime=p$, 
where $x^\prime$ is the transverse spatial position
of the single photon  
at a distance $z^\prime$ equal to the distance of the SLM from the source.
Then, the phase $e^{-i\frac{\omega t}{2} p^2}$ is imprinted by the SLM.
The map that changes the representation can be accomplished by a Fourier transform with Lohmann's type $I$ 
optical configuration such that $z^\prime=2f$, with $f$ the focal length of the spherical 
lens. The second Fourier transform with identical optical configuration maps the 
wave function back to the position representation.

\par
\subsection{Vibrational modes in trapped ions}
In order to determine Gaussian states in the vibrational modes of trapped ions \cite{Blatt2008},
one-mode conditional rotation and squeezing transformations are needed.
This constitutes the {\it first strategy} in Section \ref{Va}.

We consider a system of ions confined in an elliptical trap. 
To a good approximation, the quantized motion of each ion's centre-of-mass along the confined spatial
dimensions can be described by a quantum harmonic oscillator \cite{King1998}.
We define $\hat U_0 := \bigotimes_{j=1}^n\hat M_{{\sf R}^{(j)}}$,  
the unitary free evolution of all the harmonic motions corresponding to 
local metaplectic rotations
 $\hat M_{{\sf R}^{(j)}}$ 
in each  vibrational 
mode of the system. In this way, the Gaussian state 
to be determined can
be written in the interaction picture with respect to the free evolution as 
$\hat{\tilde{\rho}}_G:= \hat U_0^\dagger \hat \rho_G \hat U_0$, 
and the engineered metaplectic evolution (in the same representation), needed for the determination of $\hat{\tilde{\rho}}_G$, 
is written as %
$\hat U_I:=\hat U_0 \hat U = 
\exp(-i \tfrac{t}{\hbar} \hat{\sf 1}_{j-1}  \! \otimes \! \hat H_I^{(j)} \!
                                            \! \otimes \! \hat{\sf 1}_{n-j} )$. 
With this notation, the trace that appears in Eq.~(\ref{Pmenos-Pmas}) reads $\Tr[\hat{\tilde{\rho}}_G \hat U_I]$.

\par
The qubit ancilla in each single vibrational mode corresponds to two 
specific electronic states of each ion, namely $\ket{g}:=\ket{3,-}$ and $\ket{e}:=\ket{3,+}$. 
We are interested in a type of 
laser excitation in which only the motional degree-of-freedom is excited conditioned to the 
occupation of the excited level, that is, if the ion is in the state 
$\ket{g}$ nothing happens, whereas if the ion is in the state $\ket{e}$
its vibrational motion is excited. This can be accomplished with a Raman excitation of one motional sideband via the virtual excitation to an auxiliary upper electronic 
state $\ket{{\rm aux}}$, with $E_{\ket{{\rm aux}}}> E_{\ket{e}}$ \cite{Wallentowitz1999,Dalvit2006}.  The interaction Hamiltonian, in the interaction picture,
that describes the effective action of the laser over the $j^{\underline{\text{th}}}$ motional degree-of-freedom
(corresponding to a particular ion) is
\cite{Wallentowitz1999,Dalvit2006}
\beq
\label{Hamil-ions}
\hat H_I^{(j)}=\frac{1}{2}\hbar|\Omega_0|e^{i \varphi}\hat f_{k}(\hat a_j^\dagger\hat a_j,\eta)
\hat a_j^k + H.c,
\eeq
where $\hat a_j$ is the annihilation operator of the $j^{\underline{\text{th}}}$ vibrational mode considered,
$\Omega_0=|\Omega_0|e^{i\varphi}$ is the effective Raman Rabi frequency, 
$k$ corresponds to the excitation of the $k^{\underline{\text{th}}}$ upper motional sideband (blue sideband transition), $\eta$ is the Lamb-Dicke parameter, and $\hat f_{k}(\hat a_j^\dagger\hat a_j,\eta)$ is an Hermitian operator function that strongly depends on $\eta$ \cite{Wallentowitz1999}. Here we assume that each ion can be addressed individually. 
\par
The conditional rotation on the $j^{\underline{\text{th}}}$ vibrational mode occurs when the carrier sideband $k=0$ is excited and $\eta$ is not extremely small so
we have $\hat f_0(\hat a_j^\dagger\hat a_j, \eta)\approx A_0+A_1\hat a_j^\dagger\hat a_j$
\cite{Dalvit2006}. Thus, by choosing $\varphi=0$, the Hamiltonian in Eq.~(\ref{Hamil-ions}) 
can be approximated by
$\hat H_I^{(j)}\approx \hbar\omega \hat a_j^\dagger\hat a_j$, with 
$\omega=(1/2)A_1|\Omega_0|$.  
The conditional squeezing transformation of the $j^{\underline{\text{th}}}$ vibrational mode can be implemented for very small values of the
Lamb-Dicke parameter, that is, for $\eta \ll 1$, when the second blue side band is excited, 
$k=2$, so $\hat f_0(\hat a_j^\dagger\hat a_j, \eta)\approx A_0$, and 
the Hamiltonian in Eq.~(\ref{Hamil-ions}) becomes approximately
$\hat H_I^{(j)}\approx \frac{\hbar\omega}{2}(\hat a_j^{\dagger 2}e^{i\varphi}+\hat a_j^{ 2}e^{-i\varphi})$, with $\omega=A_0|\Omega_0|$.
In both cases (conditional rotation and conditional squeezing) the measurement protocol of the total phase must be initiated with the 
particular ion in the electronic state $\ket{1,+}:= (1/\sqrt{2})(\ket{e}+\ket{g})$. In addition, after the conditional evolutions, a $\pi/2$ rotation on the electronic states has to be implemented in order to obtain the probabilities in Eq.~(\ref{Pmenos-Pmas}) through the 
measurement of the population of the excited state $\ket{e}$. 
 
\begin{figure}[h]
\centering
\includegraphics[width=8cm]{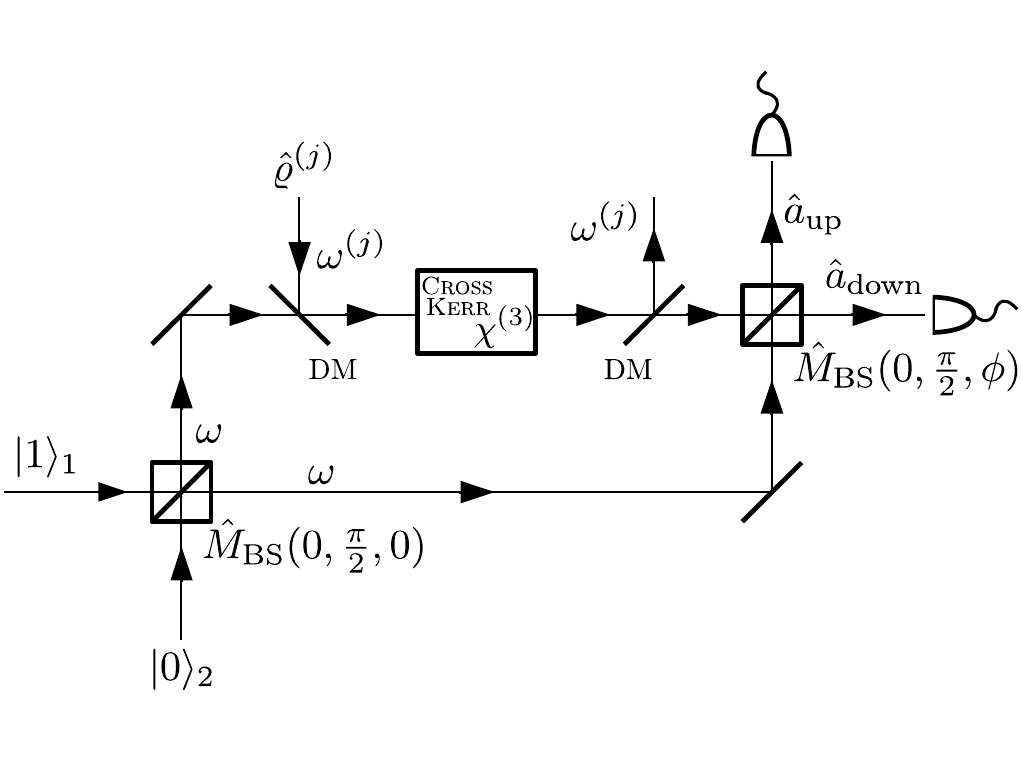}
\caption{Experimental setup of the measurement protocol of the total phase of a rotation 
of the reduced one-mode Gaussian state, $\hat \varrho^{(j)}$, corresponding to 
the $j^{\underline{\text{th}}}$ mode of a multimode Gaussian state. 
The initial qubit ancilla state  corresponds to the mode-entangled state of one photon 
(of frequency $\omega$) in the interferometer after the first beam splitter.  
The $j^{\underline{\text{th}}}$ mode of frequency $\omega^{(j)}$ is
injected and extracted from one of the arms of 
the interferometer through suitable dichroic mirrors (DM).
The rotation over $\hat \varrho^{(j)}$ is implemented by the Kerr medium
conditioned to the one-photon occupation of the upper arm or the interferometer. 
Finally, the rotation of the qubit ancilla is performed by the second beam splitter and the
measurement of 
the number of photons at the output modes determines, through Eq.~(\ref{nup-ndown}), 
the total phase $\phi_{\sf R^{(j)}}$ once we choose $\varphi=0$ and $\varphi=\pi/2$.}
\label{fig3}
\end{figure}
\par
\subsection{Quadrature modes of the quantized electromagnetic field}
In the quadrature modes of the quantized electromagnetic field highly entangled multimode Gaussian states can be generated, for example, 
in an optical frequency comb
generated by a synchronously pumped optical parametric oscillator (SPOPO) \cite{chen14,MedeirosdeAraujo2014,Gerke2015}. This is 
 the specific CV system that we will consider in this section, and the most suitable strategy to determine the Gaussian state involves the one-mode conditional rotations described in the {\it second strategy} 
in Section \ref{Va}.

In a way analogous to the CV system of vibrational modes of ions, 
here we will determine the Gaussian state 
$\hat{\tilde{\rho}}_G:= \hat U_0^\dagger \hat \rho_G \hat U_0$ 
at the output of the SPOPO crystal, in the interaction picture with 
respect to the harmonic free evolution 
$\hat U_0 := \bigotimes_{j=1}^n\hat M_{{\sf R}^{(j)}}$ 
of the fields.  
\par
\begin{figure}[h]
\centering
\includegraphics[width=8cm]{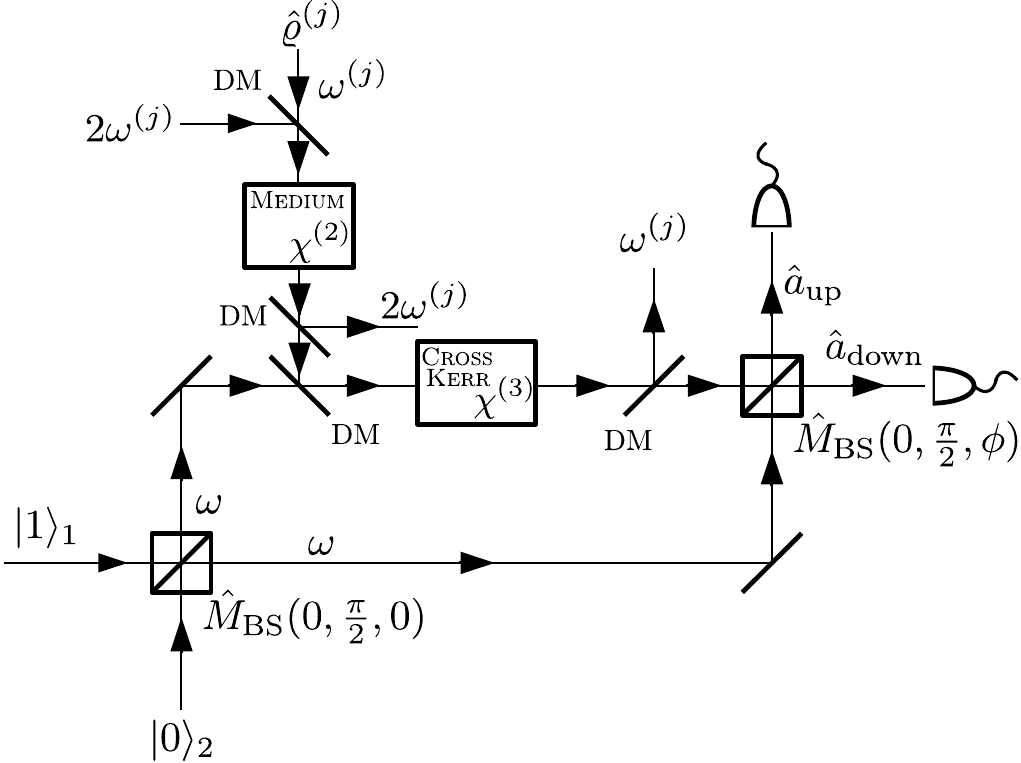}
\caption{The same experimental setup as in Fig. \ref{fig3}, but now 
before the conditional rotation in the Kerr medium a squeezing transformation 
over the state $\hat \varrho^{(j)}$ is performed.  This is done with a stimulated degenerate 
parametric down conversion in a type $I$ non-linear crystal, characterized by a 
second order electric susceptibility $\chi^{(2)}$. Thus, the down conversion process 
pumped by the field of frequency $2\omega^{(j)}$ is  stimulated by the mode of frequency $\omega^{(j)}$ 
in the quantum state $\hat \varrho^{(j)}$. The converted fields
are in the same mode $\omega^{(j)}$ producing a squeezing effect on $\hat \varrho^{(j)}$.}
\label{fig4}
\end{figure}
\par
The experimental setup of the whole measurement protocol is sketched in Fig. (\ref{fig3}).
The ancilla system is composed of one photon in the two-mode output of the first beam splitter, in the state 
\bea
\ket{\Phi}&:=&\hat M_{{\rm BS}}(0,\pi/2,0)\ket{1}_1\ket{0}_2\nonumber\\
&=&(1/\sqrt{2})(\ket{0}_1\ket{1}_2+\ket{1}_1\ket{0}_2),
\eea
where the beam splitter metaplectic operator is defined as
\beq
\label{BS}
\hat M_{\rm BS}(\psi,\theta,\phi)=e^{-i\psi\hat L_z}e^{-i\theta\hat L_y}e^{-i\phi\hat L_z},
\eeq
with $\hat L_z:=(1/2)(\hat a_1^\dagger \hat a_1-\hat a_2^\dagger \hat a_2)$, and 
$\hat L_y:=(i/2)(\hat a_1^\dagger \hat a_2-\hat a_2^\dagger \hat a_1)$.

The conditional rotation is implemented with a cross-Kerr nonlinear medium characterized by a 
third order electric susceptibility $\chi^{(3)}$, and with an interaction Hamiltonian
between the modes given by
$\hat H_K=\hbar \kappa \hat a_1^\dagger\hat a_1\hat b_j^\dagger\hat b_j $,
where $\hat a_1(\hat a_1^\dagger)$ and $\hat b_j(\hat b_j)$ are the annihilation (creation)
operators in the modes with frequencies $\omega$ and $\omega^{(j)}$, respectively.
After the interaction of the modes $1$ and $j$ at the Kerr medium
we trace out mode $j$. Then, modes $1$ and $2$  (both of frequency $\omega$) enter in a second beam splitter characterized by the metaplectic operator  
$\hat M_{\rm BS}(0,\pi/2,\phi)$.
From photo-counting measurements at the output ports of the second beam splitter
we get 
\beq
\label{nup-ndown}
\langle \hat n_{\rm up} \rangle-\langle \hat n_{\rm down} \rangle=
\Re\left[e^{-i\phi}\Tr(\hat \varrho^{(j)}\hat M_{\sf R^{(j)}})\right],
\eeq
where $\hat M_{\sf R^{(j)}}=e^{-i\theta \hat b_j^\dagger\hat b_j}$, $\theta=\kappa t$ with
$t$ the interaction time of the fields inside the Kerr medium, and $\hat \varrho^{(j)}$
the determined reduced one-mode state. By setting $\phi=0$
and $\phi=\pi/2$ we obtain the real and imaginary parts of the trace in Eq.~(\ref{nup-ndown}),
and therefore the total phase 
$\phi_{{\sf R}^{(j)}} = \arg\left[\Tr(\hat \varrho^{(j)}\hat M_{{\sf R}^{(j)}})\right]$.
It is important to notice that any value of $\theta=\kappa t >0$ serves to 
determine $\hat \varrho^{(j)}$ following the steps described in Section \ref{V}. 
Thus, the current technological limitation of very small values of the coupling constant
$\kappa$ in Kerr mediums is not a problem in our scheme. 
\par 
According to the {\it second strategy} described in Section \ref{Va}, it still remains to implement a squeezing transformation over the one-mode reduced state 
$\hat \varrho^{(j)}$ before the determination of the total phase associated with 
a rotation. This squeezing transformation over the $j^{\underline{\text{th}}}$ mode does not need to be implemented conditionally to the state of the ancilla. Therefore, we resort to the 
same experimental setup as before, but now introduce a type-$I$ non-linear crystal
characterized by a $\chi^{(2)}$ electric susceptibility before the implementation 
of the conditional rotation with the Kerr medium (see Fig. (\ref{fig4})). A pump laser beam of frequency 
$2\omega^{(j)}$  
\footnote{In the context of multimode Gaussian quantum states in an optical frequency comb, 
this laser beam can be a deviation of the laser beam that pumps synchronously the optical 
parametric oscillator inside the cavity.} 
enters, together with the field mode of frequency $\omega^{(j)}$ in the quantum state 
$\hat \varrho^{(j)}$,  in a non-linear $\chi^{(2)}$ crystal. In the approximation where the pump on the crystal is treated classically, the down-conversion Hamiltonian 
of the degenerate type-$I$ crystal is 
$\hat H\approx \frac{\hbar\omega}{2}(\hat b_j^{\dagger 2}e^{i\varphi}+\hat b_j^{ 2}e^{-i\varphi})$, which squeezes the quantum state $\hat \varrho^{(j)}$ of the 
stimulation field on the crystal. Typically, when the non-linear crystal  
characterized by $\chi^{(2)}$ is outside a cavity, the squeezing parameter $\zeta=\omega t$ (with $t$
the interaction time inside the crystal) is very small. This, however, does not represent a limitation in our protocol, since any squeezing parameter $\zeta>0$ serves 
for the determination of the Gaussian state $\hat \varrho^{(j)}$.

\section{Conclusions and final remarks}
\label{Conclusions}

We have designed an experimentally friendly method to determine Gaussian states of $n-$mode bosonic 
systems through the determination of its full covariance matrix, once the first moments of the state are experimentally determined, and local translations are implemented so as to make the state one with null mean values. In particular, we constructed three strategies to determine the one-mode reduced covariance matrices, based on the knowledge of the total phases acquired under specific one-mode metaplectic transformations that include rotations, squeezing and shears in position or momentum. 
\par
Each strategy is more suitable to be implemented in one of the three CV systems 
considered: the vibrational modes of trapped ions, the transverse spatial degrees of freedom of entangled single photons, 
and the quadrature degrees of freedom 
of $n-$mode quantized electromagnetic fields.   
Some of the one-mode transformations in each strategy must be implemented conditionally 
to the state of an ancilla qubit that, when measured, allows one to extract the total phase 
of each evolution, which bears the information of the matrix elements of the reduced single-mode covariance matrices.  
The same method used to determine the single-mode reduced covariance matrices, is applied in order to determine each pair of two-mode intermodal correlation matrices after the application of a single beam-splitter-like two-mode rotation, plus an additional single-mode rotation that does not need to be applied conditionally to an ancilla's state.  
\par
Further, the method proposed here is suitable for determining and quantifying entanglement in bipartitions having $1\times(n-1)$ modes associated with pure Gaussian states, via the measurement of only two total phases associated with two local rotations.
\par 
The strategy proposed here represents an alternative to homodyne detection in the quadrature mode CV system of the quantized electromagnetic field in which a local oscillator is not necessary, and the detector used in order to measure the one-photon qubit ancilla is a click/non-click detector.
Our strategy shows advantages in CV systems
in which the quadrature measurement is not directly accessible, as for instance in vibrational 
modes of trapped ions, and generically in networks of massive oscillators. 
In such systems the existing strategies \cite{Wallentowitz1995,Tufarelli2012} (like ours) involve the entanglement with 
a qubit ancilla, and consist in a qubit measurement from which it is possible to determine the phase-space values of the Weyl 
characteristic function of the GS. The determination of the covariance matrix in these strategies requires knowing a considerable 
number of phase-space points of the Weyl characteristic function around the origin, 
which in turn requires a lot of measurements of the qubit ancilla. In contrast, our strategy involves only a few measurements of the qubit ancilla
\par
Optomechanical systems are also CV systems 
in which the quadrature measurement is not directly accessible, and for which our strategy
could offer advantages over the existing methods \cite{Vanner2014,Moore2016}. 
In these systems the CM of the mechanical mode is indirectly determined 
by measuring the leaking field of the cavity, which is entangled with an oscillating mirror. 
It is worth noting that the determination of the CM through this method is considerable noisy. In this regard, our strategy might represent a less noisy alternative that could deserve further investigation.
\par
The method advanced in \cite{hormeyll14} to determine the CV corresponding to the spatial transverse modes of single photons can be applied in any quantum state (Gaussian or not). This strategy shares with ours the fact that it resorts to a controlled unitary 
operator implemented by a SLM, and that the 
ancilla qubit to be measured (in order to extract the second-order moments 
of the $\hat x$ and $\hat p$ operators that build the CM) is the polarization state of the photon. However, the drawback of this strategy 
is that it rests on a very precise alignment between the region in the SLM
where the unitary operation is implemented, and the region of effective 
support of the quantum state of the photon in the spatial transverse degrees of freedom. 
In contrast, our strategy is free of this alignment problem.



\acknowledgments
FN and FT acknowledge financial support from the Brazilian agencies FAPERJ, CNPq,  
CAPES and the INCT-Informa\c{c}\~ao Qu\^antica. AVH acknowledges financial 
support from DGAPA, UNAM through project PAPIIT IA101816. 
APM acknowledges the Argentinian agency
SeCyT-UNC and CONICET for financial support. We are grateful to Stephen P. Walborn,
Antonio Zelaquett Khoury, Gabriel H. Aguilar, and R. Medeiros de
Araújo for fruitful discussions.


\appendix  
 \par
\section{Total phase of metaplectic evolutions over a single-mode Gaussian state}
\label{examples}
In this Appendix we calculate the total phase acquired when different 
metaplectic evolutions of interest are performed over a single-mode Gaussian state. 
In this case the (one-mode) covariance matrix is of the form (\ref{V-one-mode}). 
For such $2\times2$ matrices, Eq.~(\ref{TotalPha3}) reduces to 
\bea                                                                
\phi_{{\sf S}}[\hat \rho_G]  &=& 
     \frac{\pi}{2} \nu^+_{{\sf S}} \! \nonumber\\
        &-&\frac{1}{2}{\rm arg}\left[ \tfrac{1}{4}\! - 
                                   \! \det({\bf V}{\bf C}_{{\sf S}})
                                   + \tfrac{i}{2} {\rm Tr}({\bf V}{\bf C}_{{\sf S}}) 
                             \right]\!\!,   \label{TotPha1M-b}                                                                                                                 
\eea
which is equivalent to the phase given in Eq.(\ref{TotPha1M}).
\subsubsection{Rotations}
\label{section:Rotations}
The unitary dynamics of a rotation is performed by the Hamiltonian of an 
harmonic oscillator $\hat H_{\sf R}=\hbar \omega (\hat a^\dagger\hat a+1/2)$, corresponding to Eq.~(\ref{Quadham}) with Hessian
${\bf H}_{\sf R}={\mathsf I}_{2}$.  

The corresponding symplectic matrix and its Cayley parametrization are
\begin{equation} \label{rot}
{\sf R} = \left(\begin{array}{cc}
                 \cos \theta & \sin \theta \\
                -\sin \theta &  \cos \theta
                \end{array} \right), \,\,\, {\bf C}_{\mathsf R} = 
{\tan}(\tfrac{\theta}{2}) \, \mathsf I_{2},    
\end{equation}
with $\theta = \omega t$. The function ${\rm Sng}\,{\bf C}_{\sf R}$ is thus 
\begin{equation}                                                                         \label{cayrot}
{\rm Sng}\,{\bf C}_{\mathsf R} = -
{\rm Sng}(\mathsf{J}
         \mathbf{C}^{^{^{\!\!\!\!{-\!1}}}}\!\!_{\mathsf R}
         \mathsf{J}) =  
\left\{ \begin{array}{rc}
        2,  &    0  < \theta <  \pi \\
       -2,  &   \pi < \theta < 2\pi \\
        2,  &  2\pi < \theta < 3\pi \\
       -2,  &  3\pi < \theta < 4\pi
       \end{array} \right. . 
\end{equation}

In order to determine the index $\nu^{+}_{\mathsf R}$ for all
$\theta$, we proceed as follows. First, according to the lines below Eq.~(\ref{CZlimit}), we fix the index $\nu^{+}_{\mathsf R}$ equal to $0$ for $\theta=0$. The continuity of the Wigner symbol (\ref{Wsmet}) in $\theta$ then allows us to put $\nu^{+}_{\mathsf R}= 0$ for $\theta \in [0,\pi)$. For $\theta = \pi$, ${\sf R}$ has eigenvalues equal to $-1$, and the symbol (\ref{Wsmet}) diverges. To surmount this difficulty we resort to the Weyl symbol (\ref{Hsmet}) before the divergence, that is, at $\theta=\pi^{-}$. 
Thus, using Eq.~(\ref{CZdef}) we write 
\begin{equation}
\nu^{-}_{\sf R(\pi^{-})} = 
\nu^{+}_{\sf R(\pi^{-})} - \tfrac{1}{2}{\rm Sng}\,{\bf C}_{\sf R(\pi^{-})} = -1({\rm mod} \,4)=3 .
\end{equation}
Due to the continuity of (\ref{Hsmet}) we have $\nu^{-}_{\sf R(\theta)} =\nu^{-}_{\sf R(\pi^{-})} $ for  $\theta \in (0,2\pi)$. For $\theta = \pi^{+}$ we employ again the Wigner symbol (\ref{Wsmet}) and use 
\begin{equation}
\nu^{+}_{\sf R(\pi^{+})} = 
\nu^{-}_{\sf R(\pi^{+})} + \tfrac{1}{2}{\rm Sng}\,{\bf C}_{\sf R(\pi^{+})} =-2({\rm mod} \,4)=2.
\end{equation}
Since (\ref{Wsmet}) is continuous in $(\pi,3\pi)$ we fix $\nu^{+}_{\sf R(\theta)} =\nu^{+}_{\sf R(\pi^{+})}$ for  $\theta \in (\pi,3\pi)$. At $\theta = 3\pi$, the Wigner symbol exhibits a second divergence, so as before we resort to the Weyl representation at $\theta = 3\pi^-$, thus getting
\begin{equation}
\nu^{-}_{\sf R(3\pi^{-})} = 
\nu^{+}_{\sf R(3\pi^{-})} - \tfrac{1}{2}{\rm Sng}\,{\bf C}_{\sf R(3\pi^{-})} = -3 ({\rm mod} 4) = 1,
\end{equation} 
and $\nu^{-}_{\sf R(\theta)} =\nu^{-}_{\sf R(3\pi^{-})} $ for $\theta \in (2\pi,4\pi)$, due to the continuity of the Weyl symbol in that interval. Finally, the Wigner symbol in the interval $\theta \in (3\pi,4\pi]$ has the index 
\begin{equation}
\nu^{+}_{\sf R(3\pi^{+})} = 
\nu^{-}_{\sf R(3\pi^{+})} + \tfrac{1}{2}{\rm Sng}\,{\bf C}_{\sf R(3\pi^{+})} =-4({\rm mod} \,4)=0.
\end{equation}
Gathering results 
we are led to 
\begin{eqnarray}
\label{nu-plus-rot}
\nu_{\mathsf R}^+ &=&  
\left\{ \begin{array}{rc}
        0,  &  \, 0  \le \theta < \pi \\
        2,  &  \,\,  \pi < \theta < 3\pi\\
        0,  &  3\pi < \theta \le 4\pi
       \end{array} \right. , \nonumber \\  
\nu^-_{\mathsf R} &=&  
\left\{ \begin{array}{rc}
        3,  &   0  < \theta < 2\pi \\
        1,  &  \!2\pi < \theta < 4\pi\\
        \end{array} \right. . 
\end{eqnarray}
We now resort to Eq.~(\ref{TotPha1M}), write $\tau=\det \bf V$$ = ab-c^2$, and get  
the result in Eq.~(\ref{TPRotA}).

\subsubsection{Squeezing}
The dynamics associated with a squeezing is now determined 
by the quadratic Hamiltonian 
\[
\hat H_{{\sf Z}_{\varphi} } = 
i\frac{\hbar \omega}{2}(\hat a^{\dag 2}e^{i\varphi} +e^{-i\varphi} \hat a^{2}), 
\]
with Hessian
\begin{equation}
\bf H_{{\sf Z}_{\varphi} } =\left(\begin{array}{cc}
                 \cos\varphi & \sin\varphi \\
                 \sin\varphi &  -\cos\varphi
                \end{array} \right).
\end{equation}  

The associated symplectic matrix and its Cayley parametrization are given, respectively, by 
\begin{eqnarray}
{\sf Z}_{\varphi} &=& \left(\begin{array}{cc}
                \cosh\zeta + \sin\varphi \sinh\zeta & -\cos\varphi \sinh\zeta \\
                -\cos\varphi \sinh\zeta &  \cosh\zeta - \sin\varphi \sinh\zeta
                \end{array} \right), \nonumber \\
{\bf C}_{{\sf Z}_{\varphi}} &=& \tanh\!\tfrac{\zeta}{2} 
               \left(\begin{array}{cc}
                \cos\varphi & \sin\varphi \\
                \sin\varphi & - \cos\varphi
                \end{array} \right), 
                \label{squeezing}  
\end{eqnarray}
with $\zeta = \omega t$. The eigenvalues of ${\bf C}_{{\sf Z}_{\varphi}}$ are $\pm \tanh\!\tfrac{\zeta}{2}$, 
hence ${\rm Sng}\,{\bf C}_{{\sf Z}_{\varphi}}=0$ and $\nu_{{\sf Z}_{\varphi}}^- = \nu_{{\sf Z}_{\varphi}}^+$. 
According to the condition (\ref{CZlimit}), we have $\nu_{{\sf Z}_{\varphi}}^+ = 0$ for 
$\zeta = \omega t=0$.
Moreover, since the symbol (\ref{Wsmet}) has no divergencies, we have
$\nu_{{\sf Z}_{\varphi}}^- = \nu_{{\sf Z}_{\varphi}}^+ = 0$ for all $\zeta$ and $\varphi$. 
Therefore, since ${\sf Z}_{\varphi}$ has positive eigenvalues, Eq.~(\ref{TotPha1M}) gives for the total phase under squeezing the result in Eq.~(\ref{TPSqA}).
%
%

\subsubsection{Coordinate Shear}
This transformation corresponds to the Hamiltonian $\hat H_{\sf F} = -\frac{\hbar\omega}{4} (\hat a^{\dag} - \hat a)^2$ with Hessian given by
\begin{equation}
\bf H_{\sf F}=\left(\begin{array}{cc}
                 0 & 0 \\
                0 &  1
                \end{array} \right).
\end{equation} 
\par
The symplectic matrix and its Cayley parametrization are, respectively,  
\begin{equation}
\label{TotCS}
\mathsf F  =  \left(\begin{array}{cc}
                    1 & s \\
                    0 & 1
                    \end{array} \right),   
{\bf C}_{\mathsf F} = \left(\begin{array}{cc}
                            0 & 0 \\
                            0 & s/2
                            \end{array} \right),   
\end{equation}
where $s=\omega t \ge 0$. 
The index $\nu_{\mathsf F}^+ $ is null by (\ref{CZlimit}), 
and the symbol in (\ref{Wsmet}) never diverges 
although the symbol in (\ref{Hsmet}) does not exist for any value of $s$. 
Thus, by Eq.~(\ref{TotPha1M}), 
the total phase 
is the result in Eq.~(\ref{TPCshA}) 
%

\subsubsection{Momentum Shear}

This transformation corresponds to the Hamiltonian $\hat H_{\sf M}= \frac{\hbar\omega}{4} (\hat a^{\dag} + \hat a)^2$ 
characterized by the following Hessian:
\begin{equation}
\bf H_{\sf M}=\left(\begin{array}{cc}
                 1 & 0 \\
                0 &  0
                \end{array} \right).
\end{equation}
\par
In this case the symplectic matrix and its Cayley parametrization are given, respectively, by 
\begin{equation}
\mathsf M  =  \left(\begin{array}{cc}
                     1 & 0 \\
                    -s & 1
                    \end{array} \right),   
{\bf C}_{\mathsf M} = \left(\begin{array}{cc}
                             s/2 & 0 \\
                             0 & 0
                            \end{array} \right),   
\end{equation}
where $ s = \omega t \ge 0$. 
By the same reasoning as in the previous example, the index $\nu_{\mathsf M}^+ $ is null,
thus, using Eq.~(\ref{TotPha1M}), we obtain the result in Eq.~(\ref{TPMshA}).
%

\bibliographystyle{apsrev}

\end{document}